\documentclass{aa}  
\usepackage{graphicx}
\usepackage{txfonts}

\newcommand{\Ms}{\ensuremath{M_{\odot}}}
\newcommand{\eg}{{\it e.g.}}
\newcommand{\cf}{{\it c.f.~}}
\newcommand{\ie}{{\it i.e.}}
\newcommand{\viz}{{\it viz.}}
\newcommand{\beq}{\begin{equation}}
\newcommand{\eeq}{\end{equation}}

\newcommand{\mg}{\ensuremath{M_g}}
\newcommand{\mphot}{\ensuremath{M_{\rm phot}}}
\newcommand{\vg}{\ensuremath{v_g}}
\newcommand{\tg}{\ensuremath{\tau_g}}
\newcommand{\td}{\ensuremath{\tau_d}}
\newcommand{\tcr}{\ensuremath{\tau_{cr}}}
\newcommand{\mcl}{\ensuremath{M_{cl}}}
\newcommand{\rh}{\ensuremath{r_h}}
\newcommand{\rc}{\ensuremath{r_c}}
\newcommand{\reff}{\ensuremath{r_{\rm eff}}}

\newcommand{\logten}{\ensuremath{\log_{10}}}
\newcommand{\kmps}{\ensuremath{\rm~km~s}^{-1}}
\newcommand{\tauh}{\ensuremath{\tau_h}}
\newcommand{\nbh}{\ensuremath{N_{\rm BH}}}
\begin{document}

\title{How can young massive clusters reach their present-day sizes?}

\author{
        Sambaran Banerjee\inst{1,2}
        \and
        Pavel Kroupa\inst{2}
        }

\institute{
           Argelander-Institut f\"ur Astronomie (AIfA),
           Auf dem H\"ugel 71, D-53121, Bonn, Germany\\
           \email{sambaran@astro.uni-bonn.de}
           \and
           Helmholtz-Instituts f\"ur Strahlen- und Kernphysik (HISKP),
           Nussallee 14-16, D-53115 Bonn, Germany
          }

\date{Received~~~~~~~~~~~~~~~; accepted~~~~~~~~~~~~~~~}

\abstract
{The classic question that how young massive star clusters attain
their shapes and sizes, as we find them today, remains to be a challenge.
Both observational and computational studies of star-forming
massive molecular gas clouds infer that massive cluster formation is primarily
triggered along the small-scale ($\lesssim0.3$ pc) filamentary substructures
within the clouds.}
{The present study is intended to investigate the possible ways in which
a filament-like-compact, massive star cluster (effective radius 0.1-0.3 pc)
can expand $\gtrsim10$ times, still remaining massive enough ($\gtrsim10^4\Ms$),
to become a young massive star cluster, as we observe today.}
{To that end, model massive clusters (of initially $10^4\Ms-10^5\Ms$) are evolved
using Sverre Aarseth's state-of-the-art N-body code {\tt NBODY7}. Apart
from the accurate calculation of two-body relaxation of the
constituent stars, these evolutionary models take into account
stellar-evolutionary mass loss and dynamical energy injection,
due to massive, tight primordial binaries and stellar-remnant
black holes and neutron stars. These calculations also include a solar-neighbourhood-like
external tidal field. All the computed clusters expand with
time, whose sizes (effective radii) are
compared with those observed for young massive clusters,
of age $\lesssim100$ Myr, in the Milky Way and
other nearby galaxies.}
{It is found that beginning from the above compact sizes, a star cluster
cannot expand by its own, \ie, due to two-body relaxation, stellar mass loss,
dynamical heating by primordial binaries and compact stars, up to the observed
sizes of young massive clusters; they always remain much more compact compared
to the observed ones.} 
{This calls for additional mechanisms that boost the expansion of a massive
cluster after its assembly. Using further N-body calculations, it is shown that
a substantial residual gas expulsion, with $\approx30$\% star formation efficiency,
can indeed swell the newborn embedded cluster adequately.
The limitations of the present calculations and their
consequences are discussed.} 


\titlerunning{Star cluster expansion}

\authorrunning{Banerjee \& Kroupa}

\maketitle


\section{Introduction}\label{intro}

Star clusters are being observed in our Milky Way and in all external galaxies with increasing detail.
However, how they form in the first place still poses a major challenge in astronomy. It
is crucial to address this fundamental question because most, if not all,
stars appear to form in embedded clusters \citep{elmg1983,lnl2003}. A key
question in this regard is how the exposed clusters' parsec-scale, smooth,
centrally pronounced, near spherical shape, observed at all ages $\gtrsim 1$ Myr,
can be explained. This
is in sharp contrast with the irregular and much extended (10s of parsecs) morphology of molecular clouds,
where stellar nurseries form through gravitational collapse and fragmentation of dense substructures. A lack
of an age range among the members of the youngest star clusters (see, \eg, \citealt{bsv2013,holly2015})
indicates that these stars form in a burst over a short period of time. This implies that
short-timescale dynamical processes, like violent relaxation \citep{spitz87}, immediately or simultaneously follow the
formation of the proto-stars, which shape the newly born cluster.

Recent unprecedented achievements of spatial and spectral resolution in IR and sub-mm wavelengths with
instruments such as \emph{Herschel} and \emph{ALMA} have revealed an intricate network of filamentary
substructures in dense regions of molecular clouds (see \citealt{andr2013} for a review). Even \emph{ALMA} 
observations of ``dark'' molecular clumps such as ``The Brick'' (G0.253+0.016) have revealed fine filamentary
structures in them that show signatures of continuing gravitational collapse \citep{rath2015}. From such
observations, it can be generally concluded that the individual filamentary overdensities and their
junctions in molecular clouds are highly compact; of typical (projected) widths of 0.1-0.3 pc \citep{andr2011,andr2013}.
Both theoretical and observational studies indicate that groups of (proto-)stars preferentially form within
the filaments and at filament junctions \citep{andr2013,schn2010,schn2012,tafhac2015,duca2011},
giving rise to $\lesssim 0.5$ pc dense protostellar clumps (or pre-cluster cloud core;
\eg, \citealt{lnl2003,tapia2011,traf2015}). Semi-analytic studies, \eg, \citet{mk2012},
also indicate similar gas clump sizes that depend weakly on their total mass. 
There is also evidence for the formation of massive star
clusters triggered by collisions between massive molecular clouds \citep{furuk2009,fukui2014,fukui2015}.

In \citet{sb2015a}, the role of violent relaxation in shaping a young cluster is
studied in detail. It is demonstrated that in order to have a smooth-profiled, spherical cluster right from
Myr-age, as seen in massive starburst clusters (see below), the stellar system involved in the aggregation
process must be ``near-monolithic''. This implies the formation of either a single monolithic proto-cluster
within a dense molecular clump
or of several sub-clusters that merge in $\lesssim1$ Myr from parsec-scale separations. In the latter case,
the exact constraints can differ from cluster to cluster; in \citet{sb2015a} the formation of the well observed
NGC 3603 young cluster is studied. Once a monolithic initial phase is attained, it's evolution can be
studied in realistic detail using direct N-body calculations, as is widely done in the literature
(see, \eg, \citealt{pketl2001,bk2007,pfkz2013,sb2014}).  
Such studies often assume a star formation efficiency (hereafter SFE), $\epsilon$, implying
the presence of primordial gas, embedding the initial stellar distribution.
The gas is to be expelled, due to feedback
from massive stars \citep{dale2015,sb2015b}, in a timescale
comparable to the cluster's dynamical time which is mimicked by a time-decaying, cluster-centric external potential.
While this approach definitely leads to oversimplification and idealization, it still captures
the primary dynamical consequences. In particular, such studies have well reproduced the detailed
observed properties of the ONC, the Pleiades, R136 and NGC 3603 young clusters \citep{pketl2001,sb2013,sb2014}.

These studies also imply that the initial mass and size of the embedded cluster and the corresponding
(clump) SFE primarily determines the subsequent evolution of the stellar system. The rapidity of
the gas expulsion is as well crucial. Both observations of embedded systems in the solar neighborhood \citep{lnl2003}
and hydrodynamical simulations of fragmentation of gas clouds and gas accretion onto proto-stars
\citep{bate2009,bate2012,bate2013},
that include radiative feedback and magnetic fields, indicate a maximum
localized SFE of $\approx30$\%. These conditions are discussed in more detail in Sec.~\ref{gasexp}.

\subsection{Young massive star clusters: observations}\label{ymsc}

Let us now turn to the star clusters of the Milky Way and other nearby galaxies. In this study,
we will primarily restrict ourselves to ``young massive star clusters'' (hereafter YMCs). A widely
accepted definition of such objects \citep{pz2010} is star clusters which have a present-day photometric mass exceeding
$\mphot\gtrsim10^4\Ms$ and are younger than $t\lesssim100$ Myr by stellar age. These limits are not
robust but are found to serve as good representative values. By stellar-evolutionary age, a 100 Myr-old stellar system
is rather old since, by this time, all of its massive stars are evolved to produce their compact remnants.
They are called ``young''
by their dynamical age, since their age is $\approx 10 - 100$ times their present-day dynamical time (crossing time
at half-mass radius), $\tcr$. Globular clusters, on the other hand, are of age
$t\gtrsim1000\tcr$, designating themselves as
dynamically old. Note that more massive and concentrated clusters will be dynamically older at a given stellar age,
since they have shorter $\tcr$. On the other hand, if the stellar system is dynamically too young,
say, $t\lesssim5\tcr$, then, phenomenologically,
it may or may not remain gravitationally bound through its future evolution. We shall designate such gas-free
dynamically young systems simply as ``association''s, as in \citet{pz2010}. The YMCs, on the other hand,
are truly gravitationally bound objects.    

A number of YMCs and associations are observed in the Milky Way, the Local Group and in other nearby
external galaxies. A comprehensive compilation of these can be found in the review by
\citet{pz2010}. In this paper, we shall refer to this compilation in order to gain insight
on the birth conditions of YMCs. This assortment of YMCs gives, in particular, a comprehensive picture of their sizes
in our and nearby galaxies. The sizes of star clusters and its implications for their birth
conditions has always been a topic of interest and debate \citep{pf2009,mk2012,pfkz2013,pf2014}.
A relatively unambiguous measure
of a star cluster's size is its effective radius, $\reff$, defined by the projected radius from its
center containing half of the cluster's total bolometric luminosity. This is essentially the cluster's half-mass
radius in projection. If the cluster is in virial equilibrium, $\reff$ is also
representative of the cluster's virial radius. A widely-used measure of the cluster's
central region is its core radius, $\rc$, of which multiple definitions are adopted in the
literature \citep{hh2003}; a commonly used one being the radial distance at
which the surface luminosity density becomes half of its central value.
The overall spatial scale of the stellar distribution is, of course, represented
by its half-mass radius or effective radius, $\reff$.

\begin{figure}
\centering
\includegraphics[width=10.0 cm,angle=0]{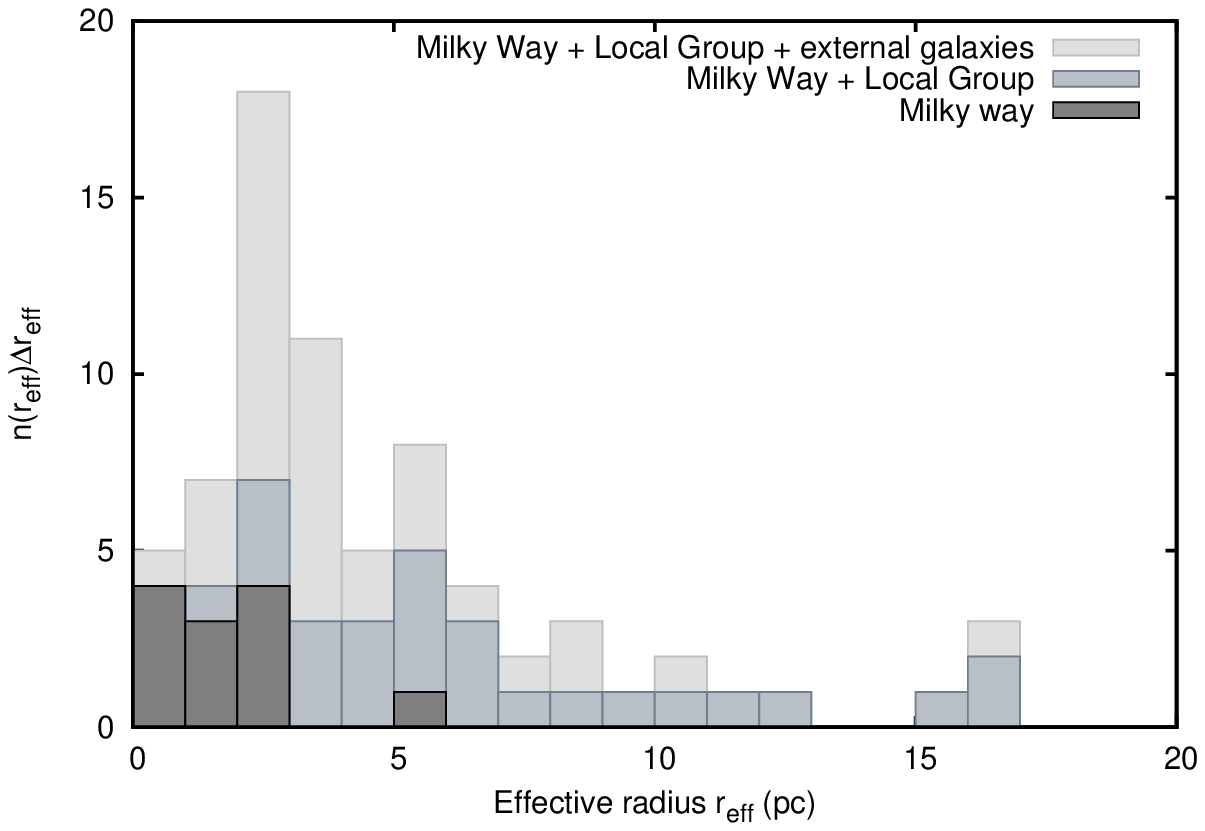}
\includegraphics[width=10.0 cm,angle=0]{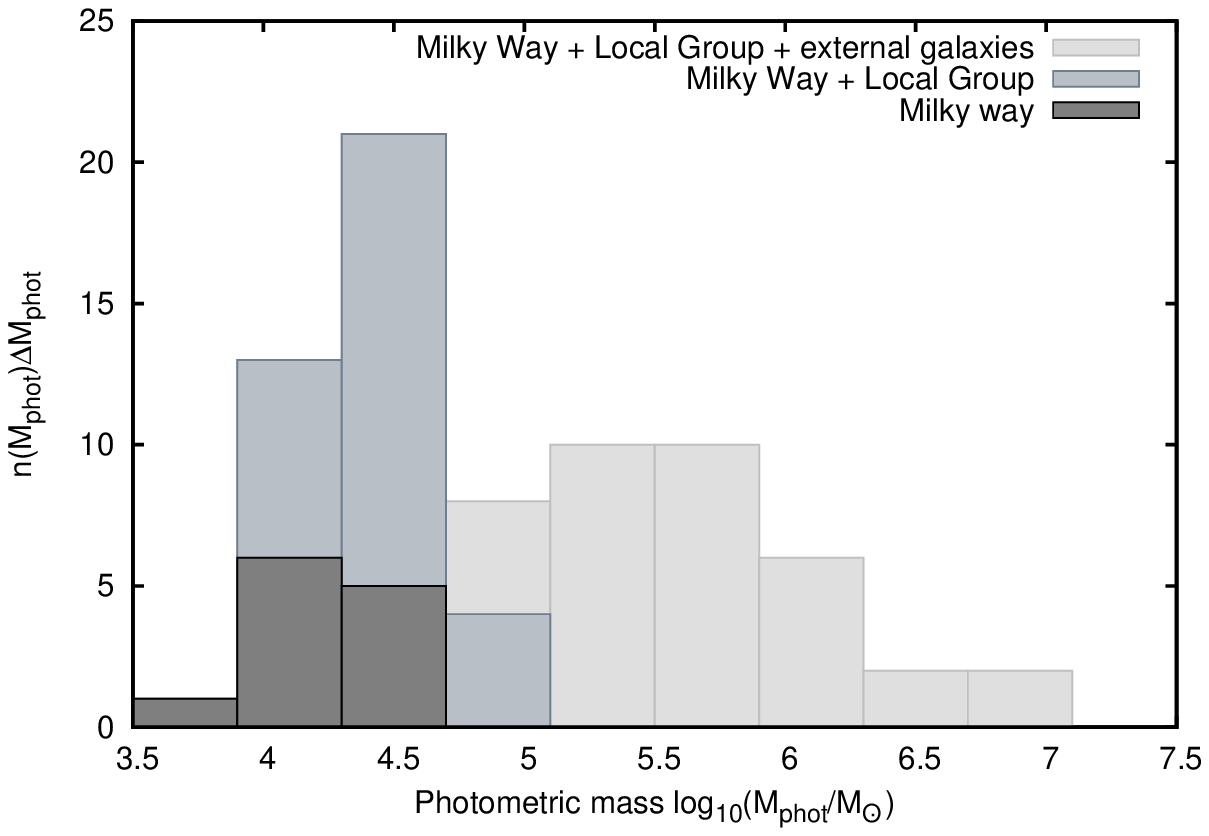}
\includegraphics[width=10.0 cm,angle=0]{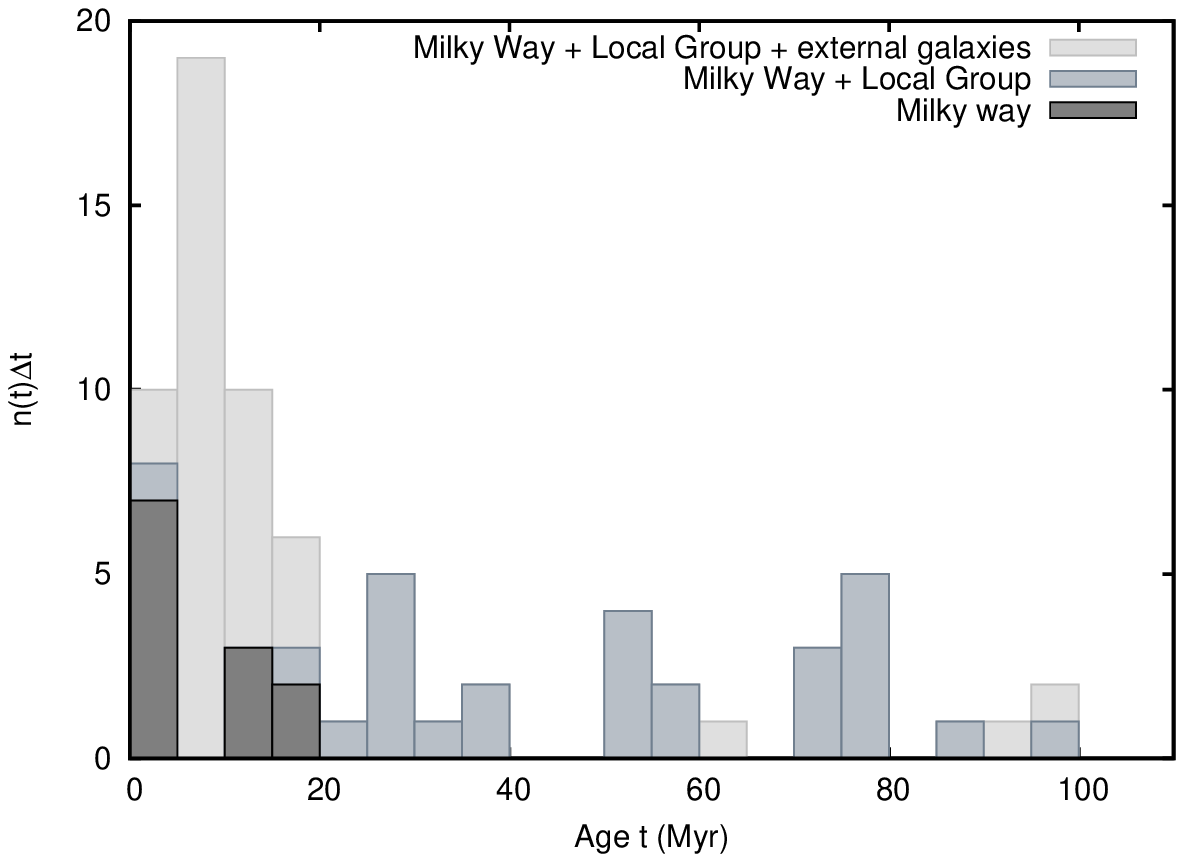}
\caption{Histograms of effective radius $\reff$ (top panel), photometric mass $\mphot$ (middle)
and stellar age $t$ (bottom) for young, massive bound star clusters (YMCs) in the Milky Way, the Local Group and in
external galaxies. The data are from Table~\ref{tab:cluster}. In each panel, the clusters from the Milky
Way, the Local Group and external galaxies are cumulatively added to construct three distributions
which are shaded differently.}
\label{fig:clusterdist}
\end{figure}

Table~\ref{tab:cluster} enlists age, $t$, total photometric mass, $\mphot$, and effective radius, $\reff$, of 
YMCs in the Milky Way (hereafter MW), the Local Group (hereafter LG) and
other nearby galaxies (hereafter OG), from \citet{pz2010}.  
Fig.~\ref{fig:clusterdist} shows the histograms of $\reff$, $\mphot$ and $t$ for the MW, LG and OG YMCs
in Table~\ref{tab:cluster}. In each panel of this figure, the distributions corresponding to the
MW members are seen to be concentrated towards lower values and are narrower. In particular, the $\mphot$
ranges are similar for MW and LG members and extends to much larger values for OG clusters.
The $\reff$ and age distributions
of LG and OG YMCs extend to much larger values than the MW ones. These trends can be explained by the
fact that clusters with larger $\reff$ survive longer in the weaker tidal fields of the LMC, SMC and other
MW satellite galaxies and the fact that the starburst OGs (NGC 4038 and NGC 4449; see Table~\ref{tab:cluster})
host much more massive YMCs (super star clusters with $\mphot\gtrsim10^5\Ms$; hereafter SSCs)
due to higher star formation activity in them.
Such SSCs potentially survive longer even in the more fluctuating and disruptive tidal fields of starburst
and merging galaxies (see, \eg, \citealt{renaud2015}).
At least some of the widest clusters are also consistent with being an ensemble
of several YMCs (see Sec.~\ref{gasexp}). A detailed discussion of the differences in the YMCs of
the MW and external galaxies is beyond the scope of this paper; see also \citet{pz2010} for a
discussion.

\begin{figure}
\centering
\includegraphics[width=10.0 cm,angle=0]{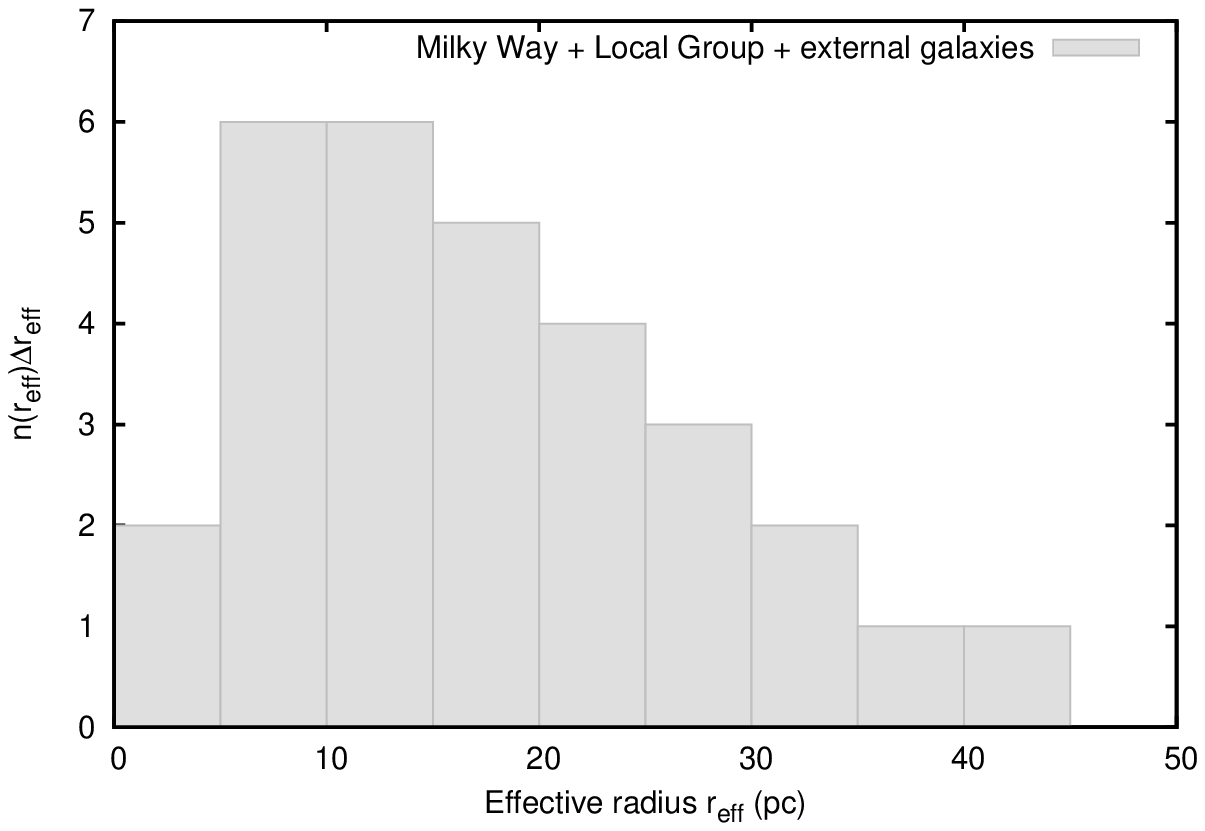}
\includegraphics[width=10.0 cm,angle=0]{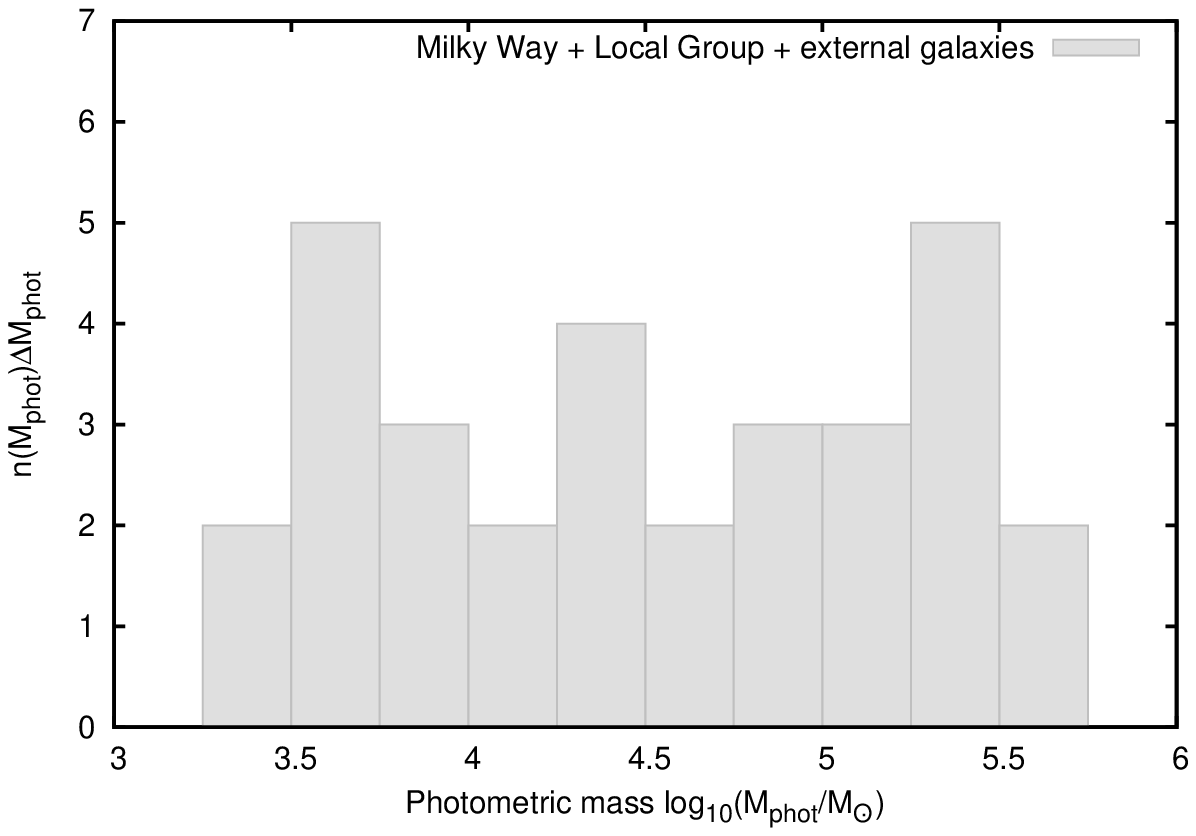}
\includegraphics[width=10.0 cm,angle=0]{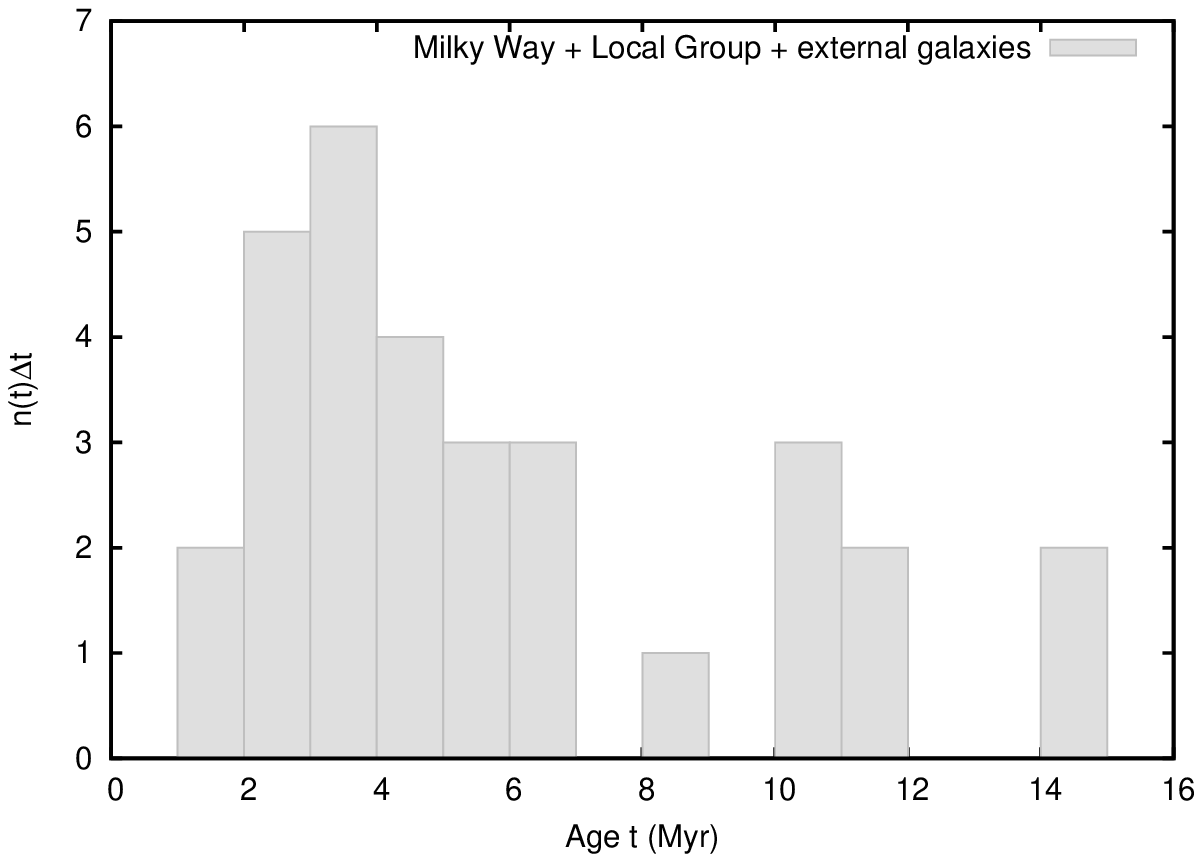}
\caption{Combined histograms of effective radius $\reff$ (top panel), photometric mass $\mphot$ (middle)
and stellar age $t$ (bottom) for young, massive stellar associations in the Milky Way, the Local Group and in
external galaxies. The data are from Table~\ref{tab:assoc}.}
\label{fig:assocdist}
\end{figure}

\begin{figure}
\centering
\includegraphics[width=14.0 cm,angle=0]{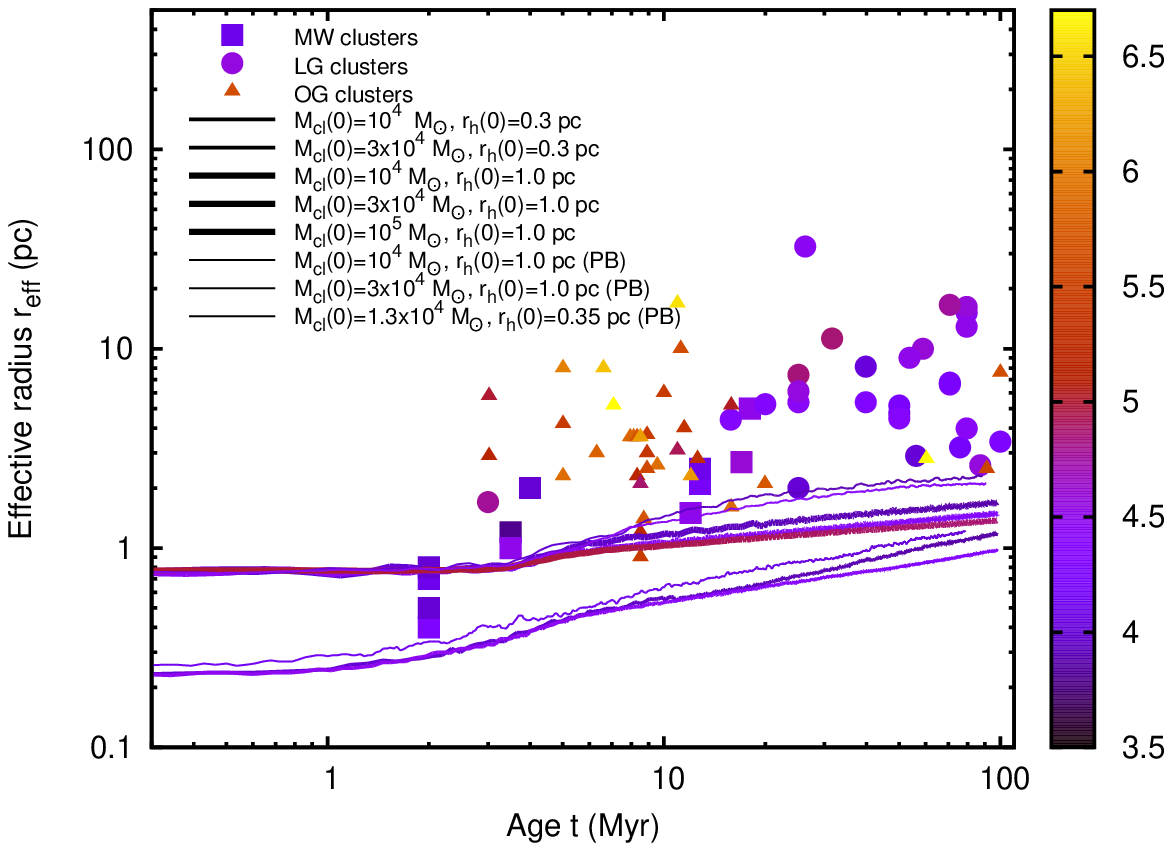}
\includegraphics[width=14.0 cm,angle=0]{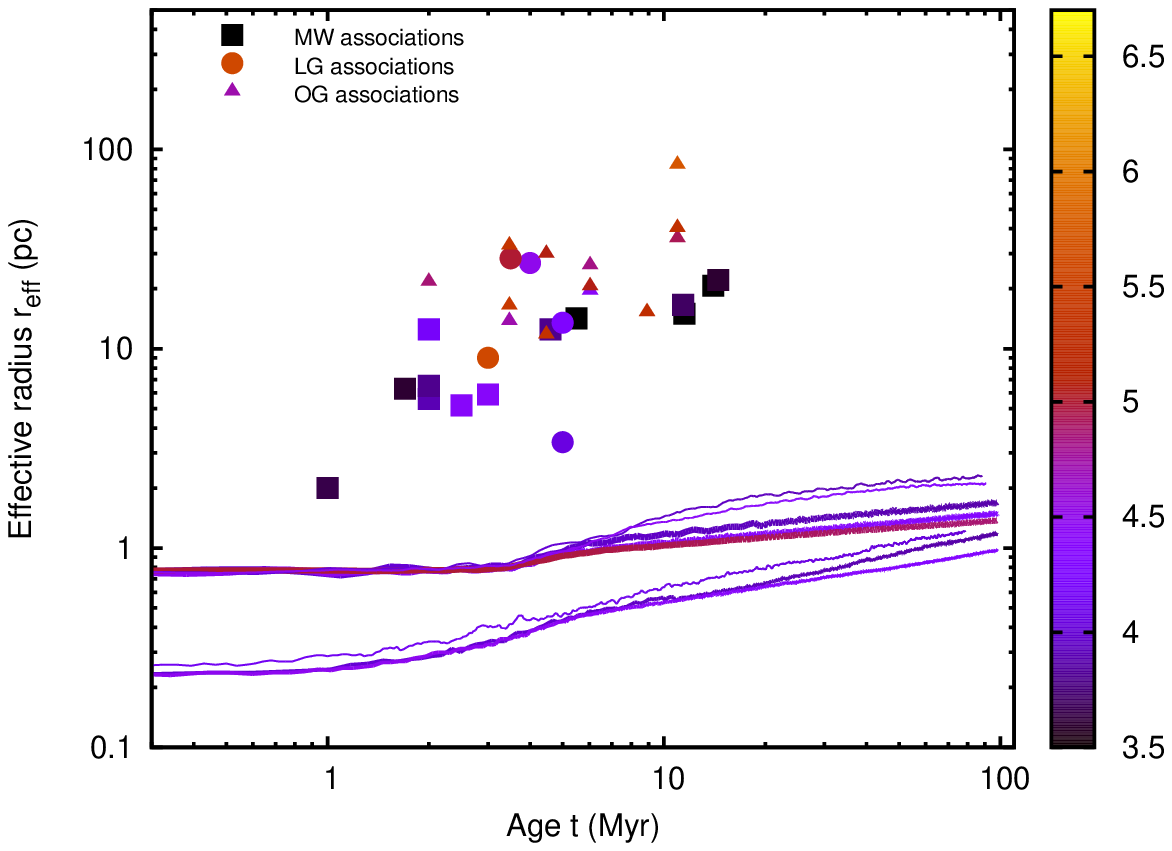}
\caption{{\bf Top}: Effective radius, $\reff$, vs. age, $t$, plot for young,
massive bound star clusters (YMCs) in the Milky Way, the Local Group and external galaxies
which are distinguished by different filled symbols. The symbols are colour-coded according
to the clusters' respective photometric mass, $\logten(\mphot/\Ms)$. These observed data are given
in Table~\ref{tab:cluster}. Overlaid in the panel are the computed curves for the
evolution of projected half-mass radius (or effective radius), $\reff(t)$, for model star
clusters with initial masses, $\mcl(0)$, and half-mass radii,
$\rh(0)$, as given in Table~\ref{tab:complist_nogas},
which do not include a residual gas expulsion phase. These curves are distinguished
according to the legends in the panel where `PB' indicates that the computed cluster
includes a realistic primordial binary population (see Sec.~\ref{nbody}).
These lines are also colour-coded
according to the corresponding clusters' instantaneous total bound mass $\logten(\mcl(t)/\Ms)$. 
As can be seen, if the clusters
evolve from compact sizes determined by substructures in molecular clouds, their
secular expansion substantially falls short of the observed sizes of YMCs (see text).
{\bf Bottom}: Here, the curves and the colour-codings are the same as above except that
the data for the young massive associations in Table~\ref{tab:assoc} are plotted.}
\label{fig:reffevol_noexp}
\end{figure}

Table~\ref{tab:assoc} enlists age, $\mphot$ and $\reff$ of the MW, LG and OG associations from \citet{pz2010}
and Fig.~\ref{fig:assocdist} shows their combined histograms. Overall, the associations are much lighter,
larger and younger compared to the YMCs (see above) which make their $\tcr$ longer and closer to their
stellar ages, thereby making them only \emph{candidates} for future bound clusters.
For the comparisons used in this study,
we will primarily consider the YMCs which are ``actual'' star clusters in the sense that a significant
fraction of their mass is gravitationally self-bound at their present age.
The filled symbols in the upper panel of Fig.~\ref{fig:reffevol_noexp}
are the YMCs from Table~\ref{tab:cluster}, in the $\reff-t$ plane,
and those in the lower panel represent this plane for the associations
from Table~\ref{tab:assoc}. These observed points and as well the computed curves (see Sec.~\ref{secexp})
in the $\reff-t$ plane are colour coded according to their
corresponding bound masses ($\mphot$ for the data points and $\mcl$ for the computed curves)
at age $t$ (the common colour scale on the right of the panels). 
We shall continue to use this data in the subsequent figures as our primary observational reference for YMCs.
In Fig.~\ref{fig:reffevol_noexp} (top panel), the $\reff$ values of the YMCs seem to follow a
moderate increasing trend with age. The youngest ones among them, of $t\lesssim4.5$ Myr, that
contain massive O-type stars, are generally referred to as ``starburst'' clusters.
The overall trend indicates a common evolutionary origin of these YMCs. 
Notably, \citet{ryon2015} present HST/WFC3 measurements of YMCs in M83 spiral galaxy using
archival data. The mean and the distribution of $\reff$s of these YMCs are
qualitatively similar to Fig.~\ref{fig:clusterdist} (top panel) and these $\reff$ values
also show an increasing trend with age.

In this work, we show how the two key observed parameters of YMCs, \viz, the effective radius, $\reff$, and the
present day mass, $\mphot$, can be used in combination to constrain the birth conditions of YMCs.
Sec.~\ref{secexp} presents N-body calculations of secular size and mass evolution of model clusters
beginning from compact initial conditions which are compared with the clusters of \citet{pz2010}.
The approach for the numerical calculations are described in Sec.~\ref{nbody}.
These computations suggest the necessity of non-secular
(\ie, by physical processes that are not inherent to the cluster)
expansion of newborn clusters and
such model calculations are presented in Sec.~\ref{gasexp}. In Sec.~\ref{himrg}, these
calculations are discussed in the context of the widely debated hierarchical merging scenario
of YMCs. Concluding remarks, based on some recent notable studies,
are presented in Sec.~\ref{conclude}.      

\section{Secular expansion of star clusters}\label{secexp}

A possible way in which a star cluster can appear within a filamentary structure of molecular gas is through a
localized but intense starburst at a privileged location in the filament, say, a at junction of
multiple filaments. Such a location is prone to reach a much higher SFE,
than that for the whole cloud of a few percent,
due to lateral contraction of the filaments
and ample gas supply through them. The nascent cluster of protostars, that would form
this way, would also be of the lateral sub-parsec scale length of the filaments, \ie, of $\lesssim0.3$ pc
(see Sec.~\ref{intro} and references therein).
A cluster with $\reff\lesssim0.3$ pc is way more compact than the presently observed YMCs, as
seen in Fig.~\ref{fig:reffevol_noexp}. The key question here is whether such a compact star cluster
can expand by its own, through its secular evolution, to attain the presently observed sizes.

To reach the present-day sizes, such a compact cluster has to expand by
a factor of $\gtrsim10$.
For a young, massive cluster, multiple physical effects contribute to its secular expansion. As long as the
massive stars remain on the main sequence (until $t\approx4.5$ Myr), mass loss due to stellar
winds is the primary driver of the expansion. The most massive O-stars remain segregated (either primordially
or dynamically) at the cluster's center (see below), so that the resulting central mass loss
due to their strong winds results in a substantial expansion of the cluster. When the most massive stars begin to
undergo supernova explosions, the central mass loss becomes more severe, causing a higher rate
of expansion. This stellar mass loss dominated expansion continues for the first $\approx50$ Myr.

After this stellar mass loss phase, the expansion of the cluster continues to be driven by
dynamical heating due to the centrally segregated black holes (hereafter BHs)
and neutron stars (hereafter NSs). Once the massive stars are reduced to BHs and NSs
by their (Type-II) supernovae and core collapse,
these remnants become the most massive members of the system,
causing them to segregate strongly and concentrate near the cluster's center.
Their high density allows to form binaries via the 3-body mechanisms \citep{hh2003}
that continue to undergo dynamical encounters.
The energy generated in encounters with the binaries (``binary heating'';
\citealt{hh1975,hh2003}) heats and expands the cluster.
The BHs, most which are of $>10\Ms$, are typically centrally concentrated within
a fraction of a parsec and are most efficient in dynamically heating the cluster
\citep{macetl2008,sb2010,sb2011,morscher2013}.
Once the majority of the BHs are depleted from the cluster through ejections due to
dynamical encounters, the NSs take over \citep{aseth2012}. The primary
uncertainty in this process is the numbers of BHs and NSs that are formed with low enough
natal kicks and are initially retained in the cluster. In the idealistic case, where
all of them are retained at birth, the BHs and NSs can contribute in dynamically expanding the
cluster for several 100 Myr.

Young star clusters are found to contain a significant fraction ($\gtrsim70$\%) of 
tight primordial binaries of massive main-sequence stars (\citealt{sev2011}, also see Sec.~\ref{nbody}).
These primordial
binaries also inject energy to the cluster and contribute to its expansion
through binary-single and binary-binary encounters and
the associated binary heating and ejections of massive
stars \citep{hh2003,sb2012,sbetl2012,oh2015}.

Finally, it should be kept in mind that any gravitationally-bound star cluster with
a realistic stellar mass distribution, can, in principle, expand purely due
to two-body relaxation (\ie, even in absence of stellar-evolutionary mass loss).
Such expansion happens in a self-similar manner (half-mass radius, $\rh\propto t^{2/3}$)
where the cluster's two-body relaxation time is a fixed multiple of the
cluster age \citep{hen1965}.
The self-similar expansion commences when the most massive particles (stars) sink to the cluster's
center via dynamical friction causing an early core collapse. This, in turn,
strongly concentrates the massive particles towards the cluster's center and they
cause the cluster to expand via dynamical heating (see above). This heating and the
expansion would, in fact, be more efficient than those due to stellar-mass
BHs (see above) since the cluster's most massive stars
are $\lesssim10$ times heavier than the remnant BHs. As demonstrated
in Figure 1 of \citet{gieles2012}, which is for low-mass clusters
of $N=256$ particles, the self-similar expansion commences only
after the $\rh(t)$ curve reaches the self-similar expansion line, $\rh(t)\propto t^{2/3}$.
For the computations here for YMCs (see Sec.~\ref{nbody}),
which are much more massive ($N\gtrsim17000$, $0.1\lesssim\rh\lesssim0.3$ pc),
the $\rh(t)$ curves would begin much above the self-similar expansion line
(\cf Figure 1 of \citealt{gieles2012}). In that case, stellar-evolutionary
mass loss would take over the cluster expansion well before the self-similar
expansion due to two-body relaxation can begin. 

\subsection{N-body computations}\label{nbody}

The dynamical relaxation of a self-gravitating many body system is most realistically
calculated using star-by-star direct N-body integration. In this work, we use the 
state-of-the-art N-body
code, {\tt NBODY7} (formerly {\tt NBODY6}; \citealt{aseth2003,aseth2012}),
to compute the secular evolution of star clusters.
In addition to integrating the individual stars' orbits using the highly accurate
fourth-order Hermite scheme and dealing with the diverging gravitational forces, during, \eg, close
encounters and hard binaries,
through two-body and many-body regularizations, {\tt NBODY7} also employs the
well-tested analytical stellar and binary evolution recipes of \citet{hur2000,hur2002}, \ie,
the {\tt SSE} and the {\tt BSE} schemes. Furthermore, mergers among stars and remnants
are included through analytic recipes. Most importantly, no softening of gravitational forces
is employed at any stage, making sure that the energetics of close encounters,
that plays a key role in the structural evolution of a star cluster, is accurately
calculated.
This numerical suite, therefore, naturally and realistically incorporates
all the physical processes that contribute to the secular evolution and expansion
of a star cluster (see above).   

To address the question of how newborn star clusters can reach their present-day sizes
from their highly compact dimensions at birth, we compute the evolution of a series of model
clusters using {\tt NBODY7}. The initial conditions of the computed model clusters are given in
Table~\ref{tab:complist_nogas}; all of them initially have Plummer profiles \citep{plum1911} 
with half mass radius, $\rh(0)\approx0.3$ pc and 1.0 pc and are
located at $R_G\approx8$ kpc Galactocentric distance (the solar distance)
and they orbit the Galactic center at $V_G\approx220\kmps$. All computed
clusters have a ``canonical'' initial mass function (IMF, \citealt{pk2001}) for the
stellar members (at zero-age main sequence; ZAMS) with
an ``optimal'' sampling of the stellar masses, \ie, the most massive star
in a cluster correlates with the initial cluster mass \citep{wk2004,wetl2013}.
The initial masses, $\mcl(0)$, of the models are taken between $10^4\Ms-10^5\Ms$.
While this mass range is appropriate for the MW and LG YMCs, the OG YMCs are 10-100 times
more massive. However, direct N-body calculations for systems over the latter mass range
is formidable by the presently available technology and one is forced to restrict until
$\approx10^5\Ms$. 
Table~\ref{tab:complist_gexp} enlists a set of initial conditions for calculations
that include the effect of
residual gas expulsion which will be discussed in Sec.~\ref{gasexp}.
For now, we consider only the calculations in  
Table~\ref{tab:complist_nogas} which are for pure stellar clusters without any
gaseous component.

Some of the computed models contain an initial primordial binary population
(see Table.~\ref{tab:complist_nogas}). While including a population of primordial binaries
makes the model more realistic, it also becomes much more compute intensive. However,
models including primordial binaries are essential to assess the role
of the latter in expanding a young cluster. In the present computed models,
we use a $f_b(0)=100$\% primordial binary fraction that follow the
``birth orbital period distribution''.
Such a primordial binary population
gives rise to the appropriate period ($P$) distribution observed for low-mass 
stellar binaries in the solar neighborhood \citep{pk1995b}.
For massive stars of $m>5\Ms$, we adopt a much tighter and narrower initial period distribution
given by a (bi-modal) \"Opik law (uniform distribution in $\log_{10} P$) that
spans over the range $0.3 < \log_{10} P < 3.5$, where $P$ is in days,
as well as ordered pairing of binary components (\ie, binary
mass ratio $\approx1$). Both conditions are motivated
by the population of O-star binaries, as observed in nearby massive clusters \citep{sev2011}.
Such observations also indicate $\gtrsim70$\% binary fraction among O-star binaries,
as consistent with the binary fraction chosen in the present calculations.
It is presently unclear at which stellar mass and how the $P$-law changes
and, therefore, in these computed models,
the discontinuous switching of the $P$-distribution at $m=5\Ms$ is chosen tacitly. 
Furthermore, primordial mass segregation is introduced in some of the initial models
with binaries (see Table.~\ref{tab:complist_nogas}), using the method of \citet{bg2008}.
The mass segregation concentrates
the massive, tight binaries in the cluster's center, maximizing the rate of
dynamical encounters they undergo and hence their heating effect (see above).        

\subsection{Secular expansion}\label{purexp}

The solid curves in the panels of Fig.~\ref{fig:reffevol_noexp} shows the computed evolution of
effective radius, $\reff(t)$, for the models in Table~\ref{tab:complist_nogas}. The same set of
curves are used for both upper and lower panels which contain the data points (filled symbols)
for observed YMCs and associations (see Sec.~\ref{ymsc}) respectively. Here, the instantaneous $\reff(t)$
is obtained by taking the arithmetic mean of the projected half-mass radii (50\% Lagrange
radius integrated over a plane) over three mutually perpendicular planes passing
through the cluster's density center. It can be seen that beginning from sizes similar
to that of the substructures in molecular clouds (see Sec.~\ref{intro}), it is impossible to attain
the observed sizes of YMCs and associations in 100 Myr. Note that realistic conditions are
used in these computed models which include stellar mass loss, retention of $\approx50$\% of the BHs and NSs formed via
supernovae and include a realistic population of tight massive primordial binaries (see Sec.~\ref{nbody}).  
For test purposes, a few computed models begin with significantly more extended size, namely, $\rh(0)\approx1.0$ pc,
which can nearly reach the sizes of the most compact observed YMCs, but still are much more
compact than most YMCs and associations. In all calculations, the cluster expands
much faster compared to its initial half-mass (two-body) relaxation
time, $\tauh(0)$ (\cf Table~\ref{tab:complist_nogas}),
since the initial expansion is driven by stellar-evolutionary mass loss (see above)
that happens in a much shorter timescale.    

\begin{figure}
\centering
\includegraphics[width=14.0 cm,angle=0]{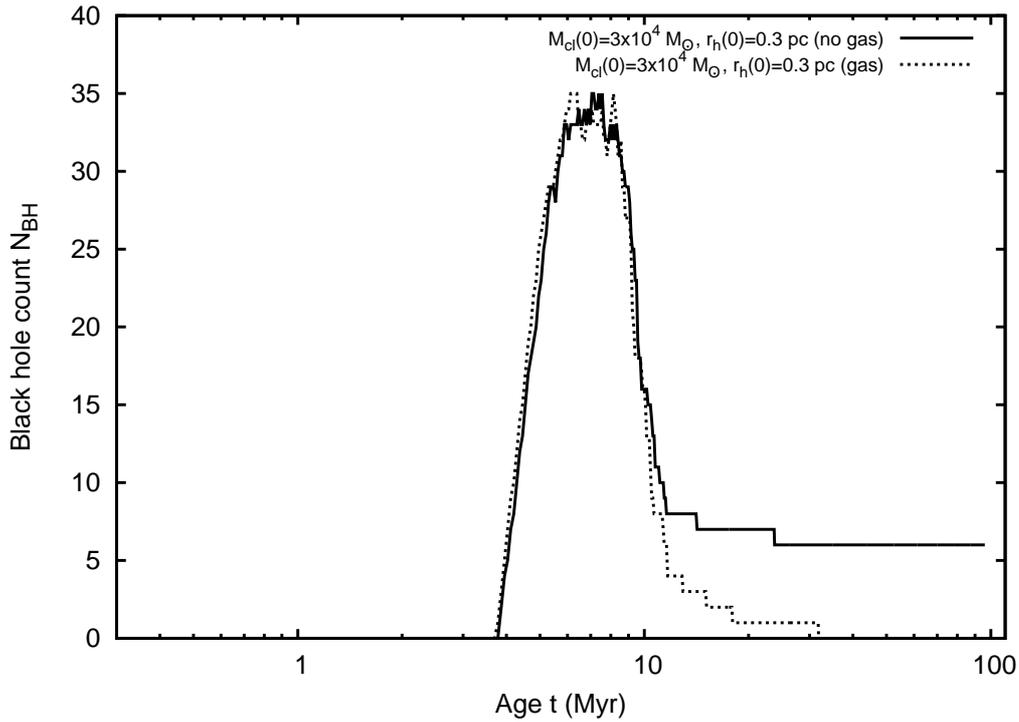}
\caption{
Evolution of the population, $\nbh$, of stellar-mass black holes (BHs) bound within the computed
clusters. The total number of BHs formed by supernovae is $\approx80$ in each case, about half
of which are retained in the cluster following their birth (by applying low natal kicks). See
text for additional detail.
}
\label{fig:bhevol}
\end{figure}

Notably, starting from the same $\rh(0)$
and for evolution time $t\lesssim10$ Myr, the computed growth of $\reff(t)$ is nearly
independent of the initial cluster mass $\mcl(0)$. As can be expected, the late-time 
growth of $\reff$ is somewhat larger for the models including primordial
binaries (Sec.~\ref{nbody}). The clusters also retain BHs and NSs that aid late-time
expansion. Fig.~\ref{fig:bhevol} (solid curve) shows the time evolution of the number of BHs, $N_{BH}$,
bound to the cluster in one of the calculations (also see Sec.~\ref{adexp}).
However, the computed values of $\reff$ always fall short substantially
of the observed ones for YMCs and associations.

Note that the computed expansion of the above model clusters are ``natural'' in the
sense that they are driven purely by dynamics and mass loss of the constituent stars.
In other words, the expansion is due to physical processes that would co-exist
in any star cluster that begins its life with a realistic zero-age stellar population.
Given that the dynamics part in the {\tt NBODY7} calculation is accurate, due to
star-by-star N-body integration avoiding force softening, the primary
uncertainties can arise from stellar evolution mass loss, in which analytic recipes
are used (see Sec.~\ref{nbody}). However, stellar evolution primarily influences the
cluster expansion during supernovae when the majority of the mass of a massive star
(ZAMS mass $>8\Ms$) is lost essentially instantaneously leaving behind an NS (typically $2-3\Ms$)     
or a BH ($10-30\Ms$ for solar metallicity). The supernova ejecta typically
exceed $10^4\kmps$ and would escape any star cluster. Hence the dominant mass
loss during the supernovae phase ($\approx 3.5-50$ Myr) effectively grosses over the uncertain
details of stellar winds during main sequence and late evolutionary stages. 
There are also uncertainties introduced from the treatments of binary
evolution undergoing mass transfer, stellar collisions and binary mergers but these have much less
impact on a cluster's expansion, compared to supernovae. 

In the present calculations, all of the stellar mass loss (wind or supernova ejecta)
are removed from the system. In reality, a fraction of the stellar wind can be slower than
the cluster's escape speed and retain in the cluster. This is also true for the
material ejected during massive stellar collisions and massive binary mergers \citep{dmink2009}.
Furthermore, gas can be externally accreted if the cluster happens to interact
with a nearby molecular cloud \citep{pflm2009}.
All of this material is either generated in-situ (\eg, for massive stellar winds and massive binary ejecta) or
eventually collected at the cluster's center as a cooling flow. As often argued,
such a gas reservoir in a star cluster can result in the formation of
second generation stars \citep{grat2012}. A quantification
of this effect is beyond the scope of this paper. However, the central gas collection
can only counteract the effect of gravitational potential dilution from stellar mass loss,
inhibiting the cluster's expansion until the first supernova. During the supernovae,
energy injected into any residual gas is sufficient to eliminate it from the cluster. 

In other words, while the details of stellar winds from massive stars
is still ambiguous, it is the supernovae which ultimately determines the
impact of the stellar mass loss
on a cluster's expansion, during its first $\approx50$ Myr. The details of the Type-II
supernova mechanism (which is largely under debate) is as well irrelevant here; the key
quantity is the total mass loss which is determined by the remnant masses. The above mass ranges
of NSs and BHs, which are adopted in {\tt NBODY7}, are supported by observations in regions
with solar-like metallicity. For lower metallicities, \eg, in the Magellanic Clouds
and several other Local-Group dwarf galaxies,
the BHs are heavier and the overall stellar mass loss is smaller. After the supernovae phase,
the primary ambiguous factor that contributes to the cluster's expansion is the
population of BHs and NSs retained at birth,
which are taken to be substantial ($\approx50$\%) in the present computations.

\begin{figure}
\centering
\includegraphics[width=14.0 cm,angle=0]{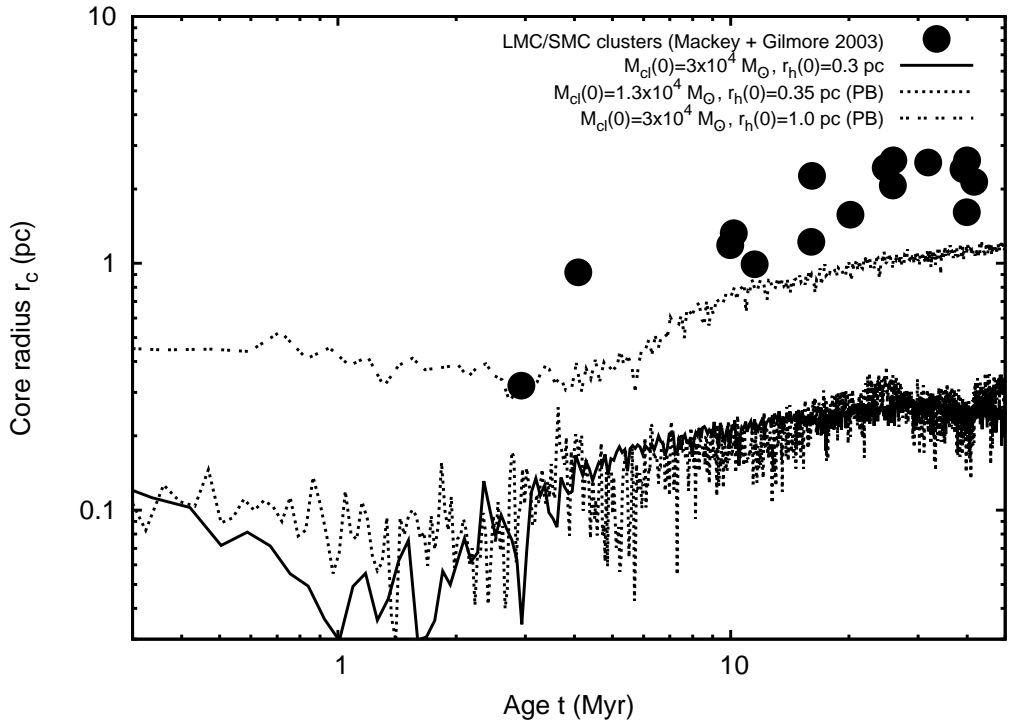}
\caption{
Evolution of core radii (solid and dashed curves), $\rc$, for representative computed clusters without a gas
expulsion phase (`PB' indicates inclusion of primordial binaries). Like the half-mass
radii, the computed values of $\rc$, starting from the filament-like compact initial conditions
(see above), substantially fall short of the observed core radii of LMC and SMC clusters
(filled circles). The latter data are obtained from \citet{macgil2003a,macgil2003b}; here, the relevant data are
extracted from the Fig.~1 of \citet{macetl2008} using the standalone {\tt DEXTER} service
({\tt http://dc.zah.uni-heidelberg.de/sdexter}).
}
\label{fig:rcexp}
\end{figure}

Fig.~\ref{fig:rcexp} shows the evolution of core radii, $\rc(t)$, for representative computed models
from Table~\ref{tab:complist_nogas}. The data points (filled circle) in this figure
are the observed core radii of LMC and SMC clusters from \citet{macgil2003a,macgil2003b}.
The computed $\rc(t)$s corresponding to the compact profiles again
well fall short of the observed values. In this context, it is important to note that
\citet{macetl2008} also computed the $\rc$ evolution for model clusters using
direct N-body calculations, which agreed reasonably with the above data. These
authors used these computations to demonstrate the dynamical heating effect of stellar mass
BHs (see above) and its effect on cluster size. However, unlike the present
calculations, those of \citet{macetl2008} start with wider $\rh(0)\gtrsim1$ pc clusters,
all the BHs are retained after formation, and the BHs are generally more massive
since lower ($0.5Z_\odot$) LMC metallicity is used. Also, unlike here,
Elson-Fall-Freeman (EFF) profiles are used as initial configurations.
These explain the differences between the present and the \citet{macetl2008} computations.
Notably, unlike $\rc$, $\reff$ approximates the virial radius of a cluster in (near)
dynamical equilibrium and hence is nearly independent of the cluster's
density profile.

The above discussion implies that although the present evolutionary calculations include
less understood physical phenomena such as massive stellar evolution, core-collapse
supernova and stellar merger, it is unlikely that a truly realistic treatment of these
would cause the clusters to expand at a much higher rate than the presently computed ones and reach the observed
sizes. This calls for additional mechanisms that would be necessary, for a newly assembled
compact cluster, for reaching the present-day observed sizes.

\section{Non-secular evolution of star clusters: primordial gas expulsion}\label{gasexp}

The computations in Sec.~\ref{purexp} do not assume any primordial gas present initially in
the cluster, \ie, the cluster is taken to be formed with 100\% local SFE. A high local SFE has been
claimed by several authors based on hydrodynamic calculations without or partially-implemented
feedback processes
(\eg, \citealt{kl1998,bate2004,kru2012,giri2012,dale2015}).
Clearly, as the computations in Sec.~\ref{purexp} show, if a young cluster is hatched along a
molecular-gas filament or at a filament junction with effectively 100\% SFE,
it cannot evolve to the presently observed sizes of YMCs and associations;
their effective radii would fall short by 1-2 orders of magnitude even after
100 Myr of evolution. There are two possible ways more extended clusters could form, \viz,
(a) residual gas dispersal and (b) mergers of less massive, compact sub-clusters.     

Observations of molecular clouds and embedded clusters suggest that in regions of
high star-formation activity, the local SFE
typically varies between a few percent to $\approx30$\%. The SFE of embedded clusters
in the solar neighborhood can also be at most $\approx30$\% \citep{lnl2003}. This is
supported by high-resolution radiation magneto-hydrodynamic (MHD) simulations of proto-star
formation \citep{mnm2012,bate2013}. Although such compute-expensive simulations are limited to the
spatial scale of a proto-star, they suggest a maximum $\approx30$\% SFE
for proto-star formation, consistently with observations. This limit occurs due
to the interplay between gas accretion onto the proto-star (sink particle) and
radiatively- and magnetically-driven outflows from the proto-star \citep{bate2013},
introducing self-regulation in the process. This fact is also supported by the
profusion of outflow activities observed from high-mass and as well
low-mass proto-stars (\eg, \citealt{bally2012,bally2015}).
The above would then imply that the maximum SFE over a region forming a population of proto-stars
is as well 30\% or less. Hence, it is realistic to assume that a proto-cluster is initially embedded
in a substantial primordial gas, the latter being subsequently cleared from the system by
the stellar feedbacks (\eg, radiation, outflows).   

A widely studied scenario that efficiently expands a cluster
from its compact initial conditions is the ``explosive''
(\ie, in a timescale comparable to the cluster's dynamical time) expulsion
of the residual gas. In this study, we utilize
the widely-used method of treating the background residual gas as an
external spherically-symmetric potential, co-centric with the cluster.
The effect of gas dispersal is mimicked by depleting this potential exponentially,
which captures the overall dynamical effect of the removal of gas (due to radiative
and mechanical feedback from the ionising O/B stars; see \citealt{sb2013,sb2015b} for
an overview). The effect of gas dispersal
on the stellar cluster depends on three primary parameters, \viz, the SFE, $\epsilon$,
the timescale, $\tg$, or the effective velocity, $\vg$ ($\tg=\rh(0)/\vg$), of
expelling gas and the
time of the commencement of the gas dispersal, $\td$. In the present calculations,
we take the representative values $\epsilon\approx33$\%, $\vg\approx10\kmps$ (the sound speed
in ionized hydrogen), $\tg\approx0.6$ Myr; see \citet{sb2013} for details. However, we also
consider the possibility of a slower rate of gas expulsion (see Sec.~\ref{slowexp}).
The relative density distribution of the stars and
gas is also crucial for the impact of gas expulsion \citep{bk2003a,bk2003b}.
Here we take both the stellar
and gas distribution to follow the same Plummer distribution, \ie, the residual gas and the
stars are ideally coupled without radial variation of $\epsilon$, maximizing
the effect of gas expulsion (also see Sec.~\ref{conclude}).

\subsection{Calculations involving ``explosive'' gas expulsion}\label{adexp}

The first part of
Table~\ref{tab:complist_gexp} lists the initial conditions for the computed models containing residual
gas which is expelled at the HII sound speed, \ie, with $\vg\approx10\kmps$. The
corresponding potential depletion timescales, $\tg$, are comparable to the clusters'
initial crossing (dynamical) times, $\tcr^\prime(0)$; see Table~\ref{tab:complist_gexp} (first part).
This makes the gas expulsion ``explosive'',
\ie, the stellar system fails to relax (via two-body interactions) and adjust with the depleting
potential, thereby beginning to expand with its own dynamical time.

\begin{figure}
\centering
\includegraphics[width=14.0 cm,angle=0]{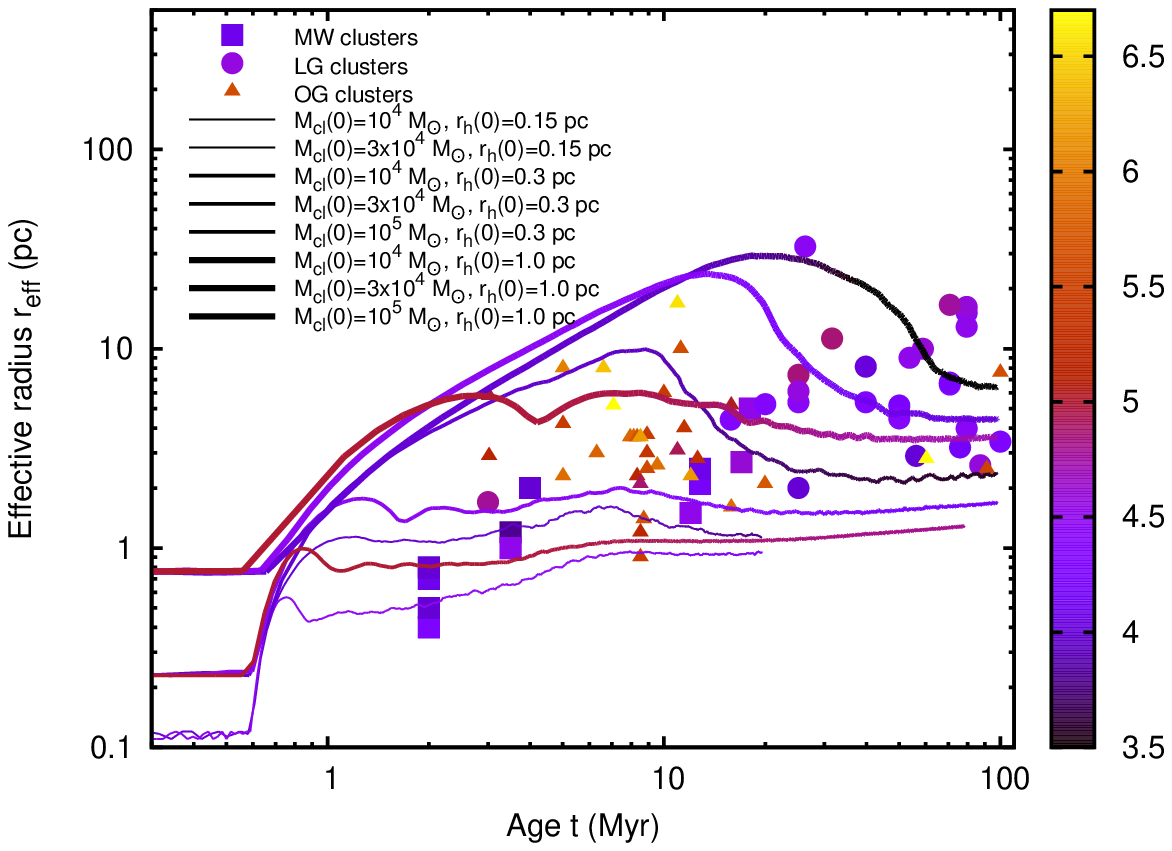}
\includegraphics[width=14.0 cm,angle=0]{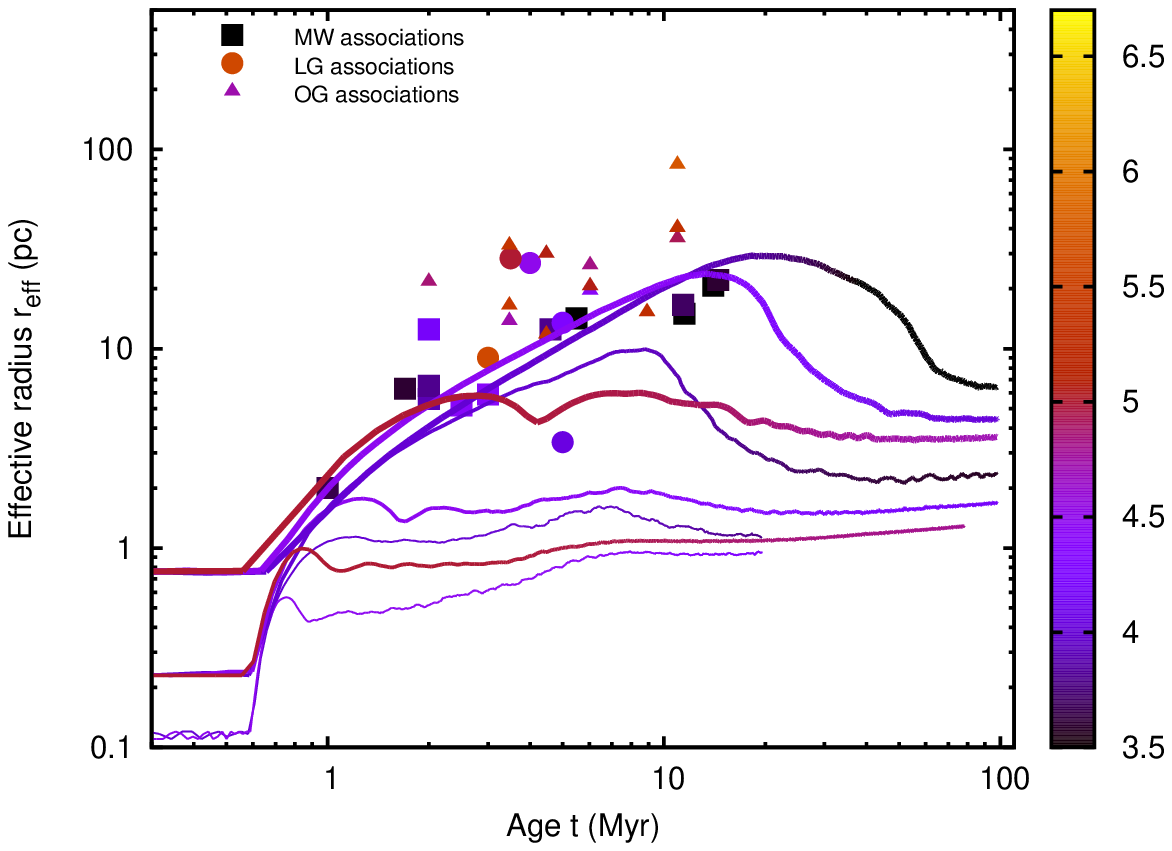}
\caption{The data points and their colour-code in both panels are
identical to Fig.~\ref{fig:reffevol_noexp}. The curves
are the computed evolution of the projected half-mass radii (or effective radii), $\reff(t)$,
including a gas dispersal phase with star formation efficiency $\epsilon\approx33$\%
(see Table~\ref{tab:complist_gexp}; top part, text for details). As in
Fig.~\ref{fig:reffevol_noexp}, the curves also follow the same colour coding according to the
corresponding clusters' instantaneous total bound mass $\logten(\mcl(t))$ (the same set of curves are
overlaid on both panels). These figures imply that even if the YMCs evolve from filament-like
compact sizes, such substantial (and explosive) gas dispersal will expand them to their
present observed sizes (half-mass radii) in the Milky way and in the Local Group (top panel).
However, to reach the sizes of the
most extended Local Group YMCs, one needs to evolve from $\rh(0)\gtrsim1$ pc half-mass radii,
unless such objects are low-mass cluster complexes \citep{bru2009}.
The latter is
also true for the young massive associations (bottom panel). See text for details.}
\label{fig:reffevol_fastexp}
\end{figure}

Fig.~\ref{fig:reffevol_fastexp} shows the $\reff$ evolutions of the above models which are
compared to the observed $\reff$ values of YMCs and associations (upper and lower panels respectively).
With an explosive gas expulsion of about 70\% by mass, the filament-like compact clusters  
can expand to reach the observed sizes of most of the MW YMCs (Fig.~\ref{fig:reffevol_fastexp}, upper panel).
On the other hand, most of the LG and OG YMCs and the associations are still a few factors
larger in size than the computed models with compact initial conditions ($\rh(0)\lesssim0.3$ pc).    
Unless these wider YMCs and associations are closely-packed ensembles of more compact and less massive
(sub-)clusters, the corresponding $\reff$ values can only be arrived from more diffuse initial configuration
with $\rh(0)\approx1$ pc, as seen in Fig.~\ref{fig:reffevol_fastexp}. 
On the other hand, $\rh(0)\approx1$ pc clusters lose more stars and become lighter in
their expanded phase, than their more compact counterparts (see color coding of the
curves in Fig.~\ref{fig:reffevol_fastexp}).

\begin{figure}
\centering
\includegraphics[width=14.0 cm,angle=0]{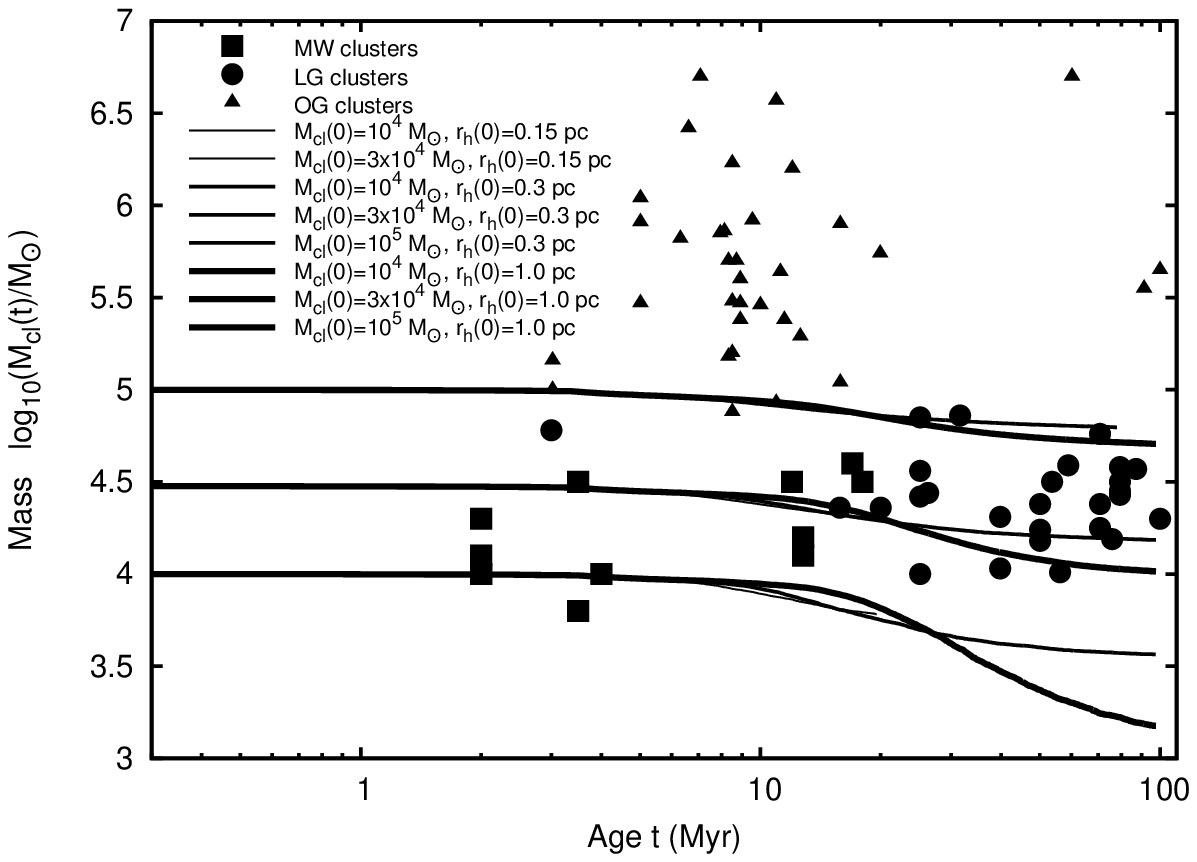}
\includegraphics[width=14.0 cm,angle=0]{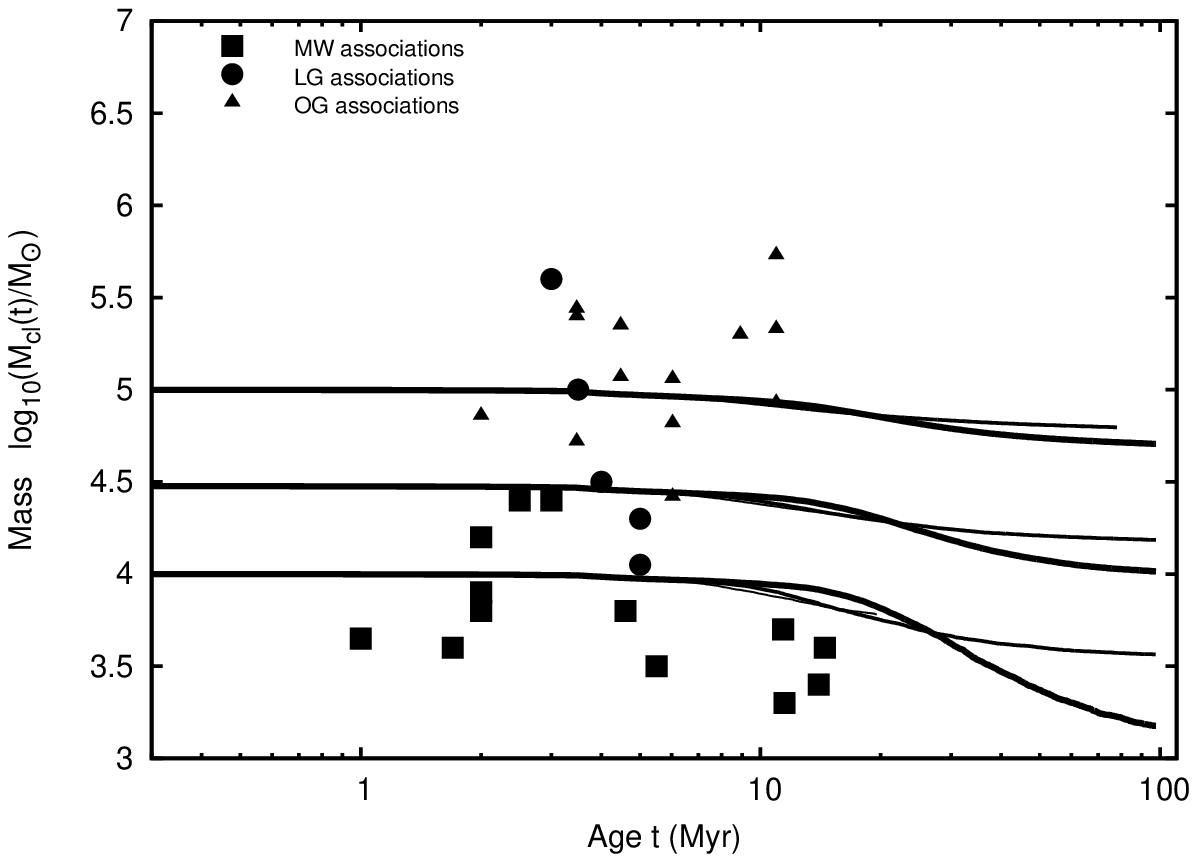}
\caption{The curves give the evolution of the instantaneous total bound stellar mass,
$\logten(\mcl(t))$, for the calculations in
Fig.~\ref{fig:reffevol_fastexp} that include an initial gas dispersal phase
(see Table~\ref{tab:complist_gexp}; top part, text). In both
panels, data points corresponding to the same young massive clusters and associations (in top and bottom panels
respectively) as in Figs.~\ref{fig:reffevol_noexp} and \ref{fig:reffevol_fastexp} are used, but without
colour coding. The same set of curves are overlaid on both panels and the legends are given in the top
panel.}
\label{fig:mphotevol_fastexp}
\end{figure}

Fig.~\ref{fig:mphotevol_fastexp} shows
the evolution of the bound mass, $\mcl(t)$, for the computed models in Table~\ref{tab:complist_gexp}
(first part), which are compared with the photometric masses, $\mphot$, of the observed YMCs and
associations. It shows that so far only the clusters' masses are concerned, the computed initial mass range of
$10^4\Ms<\mcl(0)<10^5\Ms$ reasonably covers the $\mphot-t$ space of the MW and LG YMCs and associations.
The OG sample comprises 10-100 times more massive systems (see Sec.~\ref{ymsc})
and hence similarly larger values of $\mcl(0)$ are required to cover these masses, which is beyond the
present scope (see Sec.~\ref{nbody}). However, it can be said that such massive clusters
would lose much less mass compared to their less massive (computed) counterparts (see below).

From Fig.~\ref{fig:reffevol_fastexp}, a general trend can be recovered, \viz, beginning from
a given $\rh(0)$ and for fixed $\epsilon$ and $\vg$, a cluster with larger $\mcl(0)$    
loses less mass and as well expands less. This trend is what can be expected since for a
fixed $\rh(0)$, larger $\mcl(0)$ implies a larger binding energy for the system. Also, their higher
stellar density results in more efficient violent relaxation causing a larger fraction of
stars to be retained during the expansion due to gas expulsion. This trend, however,
makes it prohibitive to achieve the observed $\reff$s of most of the extended YMCs
($\reff\gtrsim1$ pc) starting from the compact initial conditions,
without compromising the mass, as Fig.~\ref{fig:mphotevol_fastexp} shows.

Starting from $\rh(0)=0.3$ pc, the computed cluster with $\mcl(0)\approx10^4\Ms$ reaches
the sizes of most MW YMCs while still more compact than most of the LG VYMCs, but its
mass substantially falls short of the VYMC masses (\cf Fig.~\ref{fig:mphotevol_fastexp}).
The more massive $\rh(0)=0.3$ pc
clusters maintain VYMC-like masses but substantially fall short of the observed sizes
of most VYMCs. While the OG clusters exceed the computed mass range, it is clear that
the size discrepancy would only increase for them. On the other hand,
with the $\rh(0)\approx1$ pc models, only the largest $\reff$s are reachable, but again
compromising $\mcl$ (see Figs.~\ref{fig:reffevol_fastexp} \& \ref{fig:mphotevol_fastexp}).
The same holds if one decreases the SFE instead. The above discussion implies that
while the most compact MW, LG (and OG) YMCs are naturally achievable from compact
filament-sized initial clusters via residual gas expulsion,
in terms of their masses and sizes, the same seem to require fine tuning
for the more extended VYMCs (and the associations).

In passing, Fig.~\ref{fig:bhevol} shows the evolution of $N_{BH}$ for a computed model
with explosive gas expulsion (dotted curve). The lesser retention of BHs, in this case,
is likely due to larger expansion and hence lower central density of the cluster.     
The lower central stellar density largely reduces the dynamical friction
on the mass-segregated BHs comprising the central BH sub-cluster (or the
``black core''; \citealt{macetl2008}). This causes them to effectively act like an
isolated small-$N$ cluster, having short two-body relaxation time, hence
they evaporate rapidly. The effect of a star cluster's birth condition on its
compact-star population is itself an emerging topic with wide implications,
which will be presented elsewhere.  

\subsection{``Placid'' or ``slow'' gas dispersal and its implications}\label{slowexp}

Contrary to (near) explosive gas expulsion, a ``placid''
mode of gas expulsion also similarly
expands a cluster but causes much less mass loss in stars.
In this case, the gas is
dispersed over a timescale that is much longer than the cluster's crossing (dynamical) time.
This allows the cluster to adjust itself with the changing potential and expand
while remaining in near dynamical equilibrium, \ie, evolve in a quasi-static (or adiabatic; \citealt{pk2008})
fashion.
The timescale of expansion is, of course, determined by the gas dispersal time $\tg>>\tcr(0)$.
The cluster, therefore, expands much slower as opposed to the abrupt expansion at $\td$
for the explosive case (see Sec.~\ref{adexp}).   

A motivation for considering the possibility of such ``slow'' gas dispersal is several observed
gas-embedded regions containing massive O/B-stars which are several Myr old. Perhaps the
most well studied example is the Galactic embedded region W3 Main that is continuing to
form stars for at least 3 Myr \citep{fglsn2008,bik2014}.
Another more recently cited example is the W33 complex \citep{messi2015}. 
In such regions, while there are individual ionized bubbles around each massive star,
the overall stellar system still seems to be deeply embedded in molecular gas. A common
feature of such embedded stellar populations is that the stars are distributed
in a sub-structured fashion over significantly larger spatial scales than (near) spherical
clusters; typically over a few parsecs to 10s of parsecs. While, at present, it is
unclear why such systems fail to clear the gas early, unlike the starburst clusters,
the wider distribution of the ionizing O/B stars can be a clue. As argued in \citet{sb2015b},
such a condition might lead to evacuation of gas only around individual OB stars and
not in bulk.
In that case, the overall gas is dispersed in a much longer timescale than dictated
by $\vg\approx10\kmps$ which is also much longer compared to the stellar system's dynamical time
(\ie, placidly).

\begin{figure}
\centering
\includegraphics[width=14.0 cm,angle=0]{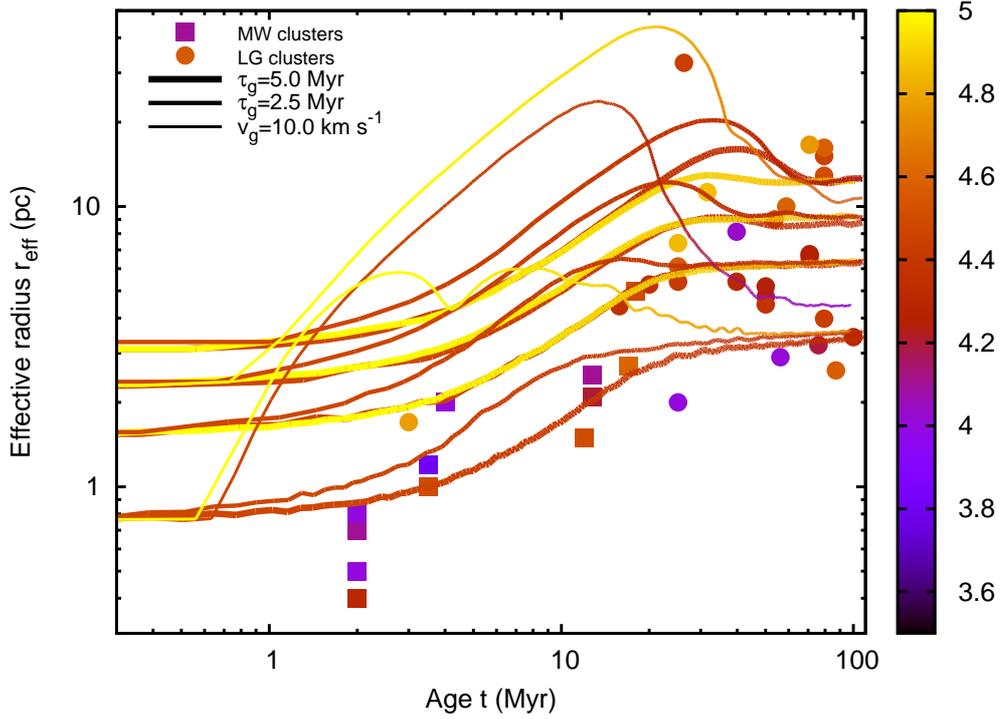}
\caption{Curves representing the evolution of projected half-mass radii (or effective radii), $\reff(t)$,
for computed model clusters that include a gas expulsion phase. Unlike those in the previous figures,
these computed clusters initiate from much more extended stellar distributions with initial half-mass radius,
$\rh(0)>1$ pc (see Table~\ref{tab:complist_gexp}; bottom part). Furthermore, the gas expulsion
timescales, $\tg$, are of much longer values as
indicated in the legend, making the gas expulsion `placid’ in these cases (except for
$\vg=10\kmps$ which causes explosive or explosive expulsion as in the previous cases). For
$\epsilon\approx30$\% star formation efficiency, these extended model clusters expand to
the observed sizes of the most extended young massive clusters in the Local Group (the filled
symbols) while also covering the mass ranges appropriate for them. Note that with slow or placid
gas expulsion, the clusters tend to reach sizes that are nearly independent of their initial masses, unlike
their explosive counterparts.
As for the previous figures, the symbols and the curves are colour-coded according to the corresponding current
total (photometric) mass.}
\label{fig:reffevol_slowexp}
\end{figure}

\begin{figure}
\vspace{-0.2 in}
\centering
\includegraphics[width=10.0 cm,angle=0]{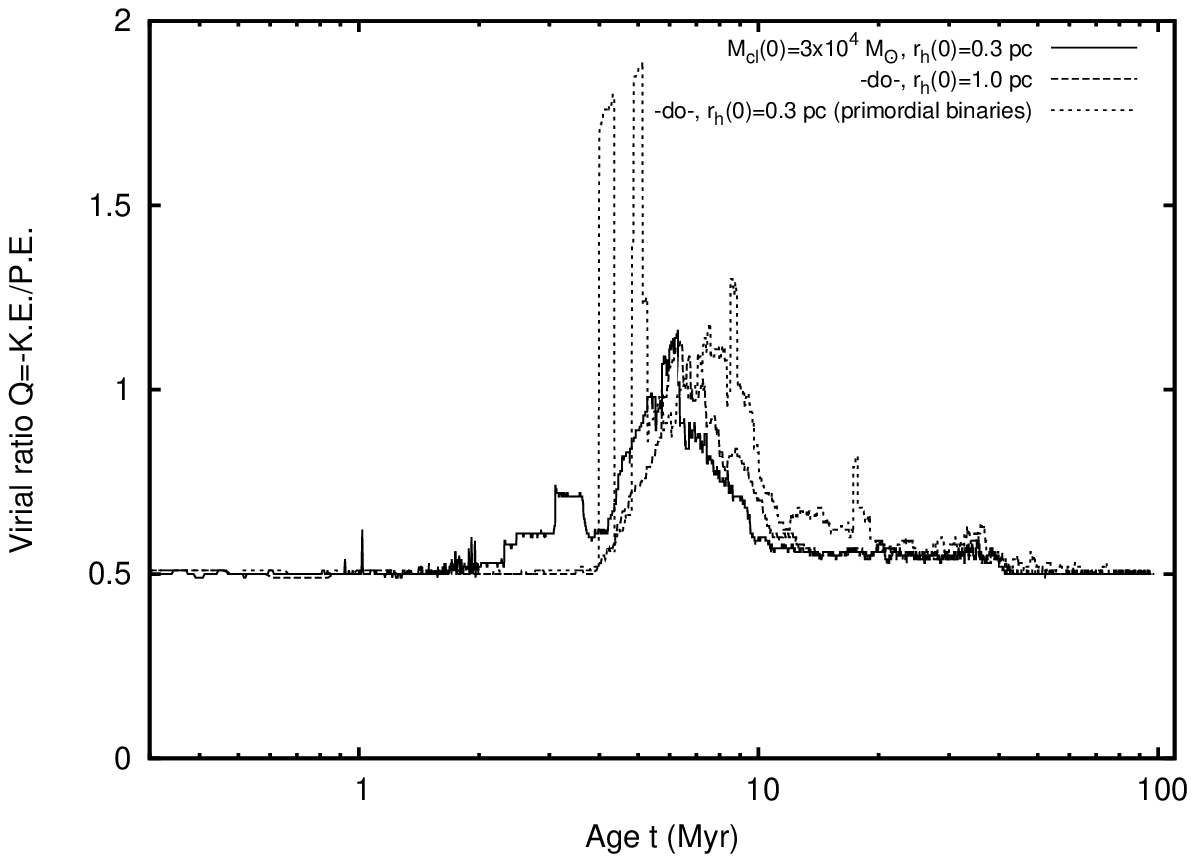}
\includegraphics[width=10.0 cm,angle=0]{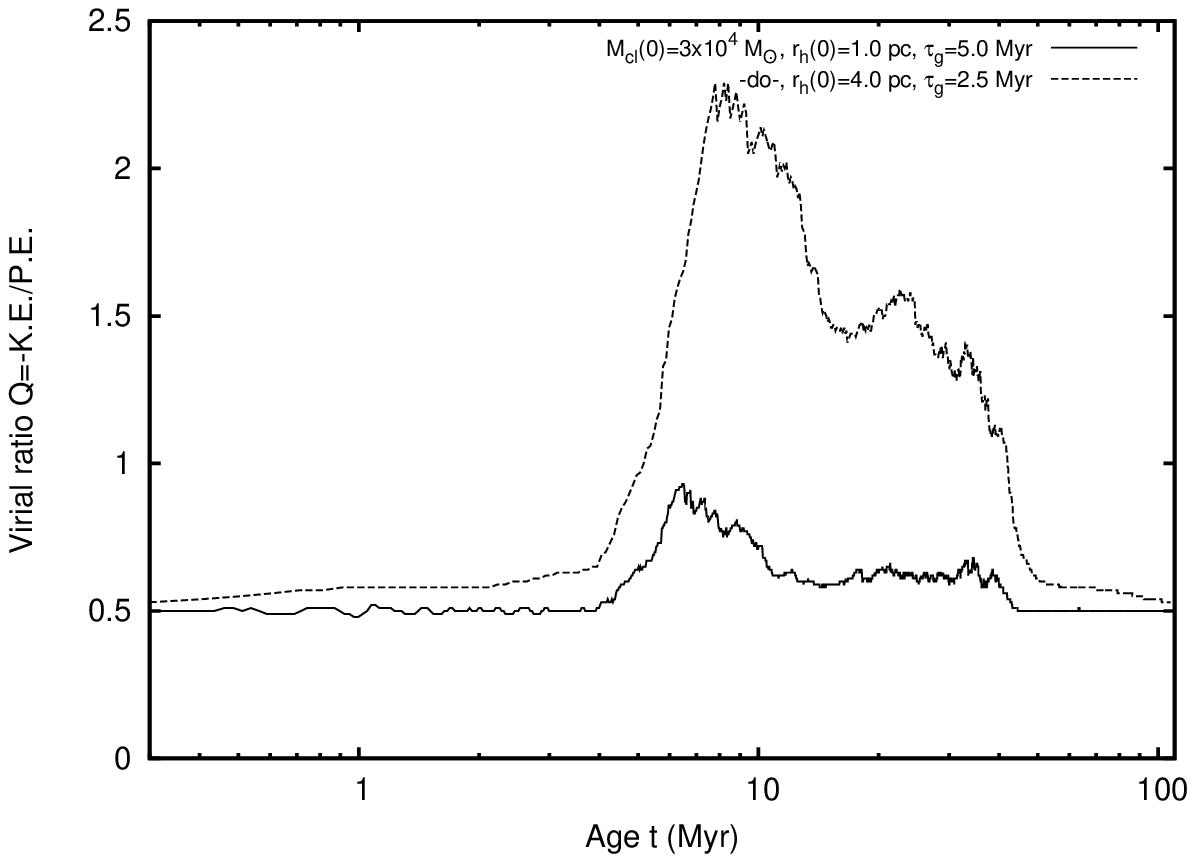}
\includegraphics[width=10.0 cm,angle=0]{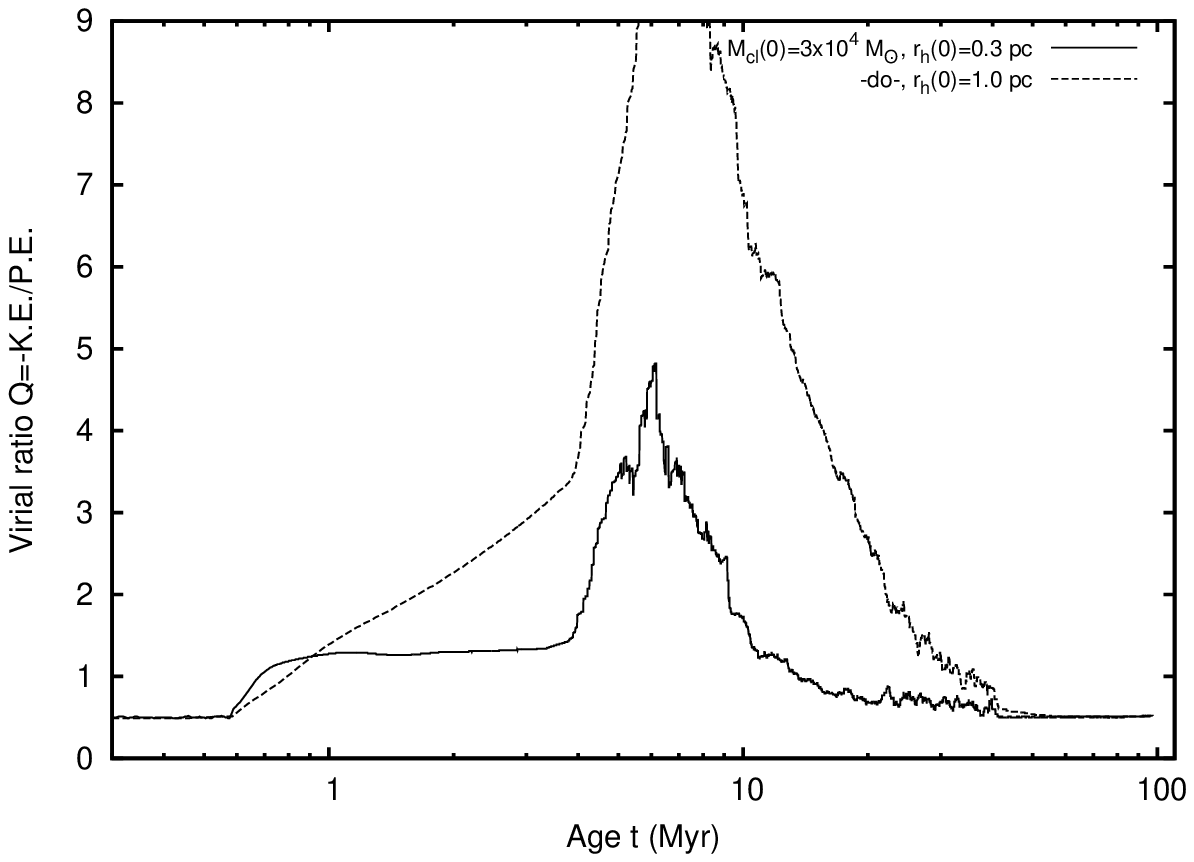}
\caption{
Evolution of the virial ratio, $Q$, \ie, the ratio of the total kinetic to potential energy
of the bound cluster members for representative computed cluster models
without gas expulsion (top panel) and those with placid
(middle) and explosive (bottom) gas expulsion. For the cases without and with placid gas expulsion,
$Q\approx0.5$, \ie, the system is in dynamical equilibrium, over most of the evolutionary times. In these
two cases, $Q$ temporarily grows (\ie, the cluster becomes supervirial)
after $t\approx4$ Myr (that corresponds to a rapid expansion of the clusters' core)
due to substantial mass loss due to supernovae. This phase lasts $\approx50$ Myr over which the BHs and the NSs form.
The (placid) gas expulsion aids this non-dynamical-equilibrium expansion phase of the cluster resulting
in larger growth of $Q$ (\cf, middle and top panel). For explosive gas removal, a cluster expands
in a super-virial manner from the beginning of gas expulsion ($\td\approx0.6$ Myr) and becomes
even more super-virial when the supernovae begin (bottom panel). Like the other cases, the system eventually
returns to dynamical equilibrium when the supernova mass loss is quenched. As expected, in all the cases, the growth of
$Q$ is larger for the initially more extended cluster.   
}
\label{fig:Qevol}
\end{figure}

The second part of Table~\ref{tab:complist_gexp} lists the initial conditions for
computed models where longer $\tg=2.5$ Myr and 5 Myr are considered. The values of $\mcl(0)$
are similar to the previously computed models but much wider $1.0{\rm~pc}<\rh(0)<4.0{\rm~pc}$
are considered. For simplicity, we consider only initially spherical systems without
any substructures, since the objective is to study the overall mass and size
evolution of the system. Fig.~\ref{fig:reffevol_slowexp} shows the corresponding computed
$\reff(t)$s. A few additionally computed models with $\vg\approx10\kmps$ are shown
for comparison. As expected, $\mcl$ does not decline significantly during the
evolution. Most of the stellar mass loss is due to the supernovae ejecta and dynamically ejected
stars which are much less than that due to stars becoming unbound by explosive gas
expulsion. Interestingly, unlike the case of explosive expulsion, placid expulsion
causes clusters of different $\mcl(0)$ to approach nearly the same $\reff$, for a given
$\rh(0)$ (and $\epsilon$). This is also due to the fact that the associated stellar mass loss being
small compared to the total bound mass, the final cluster size is essentially determined by
the relative depletion of the background gas potential. This is as expected from adiabatic
cluster evolution \citep{pk2008}. Fig.~\ref{fig:Qevol} compares the
evolution of the virial ratio, $Q$, between different modes of cluster expansion; see
caption for explanation. The above property implies that in order to achieve an
observed cluster size through slow gas expulsion, one only needs to set the appropriate
initial size, \ie, fine tuning is hardly necessary.     

The key drawback of explaining the extended YMCs (or any cluster in general)
as an outcome of slow gas dispersal is the fact that the depletion process is unlikely
to remain slow (or placid) for its entire duration. When the Type-II supernovae
commence (from $\approx4$ Myr age),
their energy input into the gas could be sufficient to the eject
the gas explosively. If this happens, then, for the extended
initial clusters used in Fig.~\ref{fig:reffevol_slowexp},
the rapid gas expulsion (together with the tidal field) would largely dissolve them.
Hence, although pedagogically interesting, slow gas expulsion is unlikely to play
a role in bound cluster formation with age $\gtrsim 4$ Myr. Hence, it seems more
likely that the few - 10s pc YMCs are ensembles of closely-located YMCs
($\sim10$s of them), forming low-mass cluster complexes, thus being
younger, low-mass versions of ``faint-fuzzy''
objects \citep{bru2009,ams2014}. Future observations of the light profiles of these objects
would help to better understand their nature.

\section{Cluster formation through hierarchical mergers}\label{himrg}

So far we have considered monolithic or in-situ formation of star clusters. 
Due to the substructured and filamentary conditions in molecular clouds in which stars form
(see Sec.~\ref{intro}), it is plausible that YMCs may also arise due to sequential
mergers of less massive sub-clusters \citep{longm2014,fuji2012,sm2013,sb2015a}.
These sub-clusters, where
the massive proto-stars preferably form and which would appear along
the filaments and at filament junctions, would fall in 
through the molecular cloud into the deepest part of the potential well where they
would merge into a single proto-cluster. This overall picture is revealed
in several recent hydrodynamic simulations of ``cluster-sized'' molecular clouds
that include radiative stellar feedback.

In this scenario, the two following
cases can be distinguished, namely, (a) a star-burst
occurs over the free-fall time of the cloud so that the newly formed proto-stellar
sub-clusters, contributing to the formation of the YMC, and the gas fall in concurrently
(see, \eg, \citealt{bate2004,bate2009}) and
(b) the star-formation rate is diminished (\eg, due to stellar feedback and heating
of the gas as in \citealt{dale2015}),
so that the ensemble of the contributing proto-sub-clusters form
over a timescale substantially longer than the free-fall time. Traditionally, case `a' is referred
to as ``coupled'' star formation and case `b' as ``decoupled''. Coupled star formation would
essentially lead to a single, monolithic embedded cluster after the merger phase,
as studied
in Sec.~\ref{gasexp}. On the other hand, decoupled star formation implies cluster
assembling independent of the motion of the gas and over timescales longer than
the free fall time.

In any case, if a cluster has to form and evolve from a young age (a few Myr
like the youngest Galactic YMCs), the sub-clusters
should fall in from sufficiently close separation so that they can merge
early enough. This is demonstrated in \citet{sb2015a} for the case of the
$\approx1$ Myr old NGC 3603 young cluster, for which the sub-clusters must
merge from $\lesssim2$ pc. This implies that although star formation is
often found to occur over $\gtrsim10$ pc regions, only a part of the newly formed stellar
structure can actually comprise a YMC. There is mounting observational
evidence of young stellar and/or proto-stellar sub-clusters  
packed within pc-scale regions (\eg, \citealt{massi2014,rz2015}).
Furthermore, as demonstrated in \citet{sb2015a},
the ``prompt'' merger of several gas-filament-like compact sub-clusters produces a
similarly compact cluster. Hence, as the calculations in Sec.~\ref{purexp} suggest,
the presently observed YMC sizes would be unreachable for the newly assembled cluster
via pure secular expansion. This is as well true for the individual sub-clusters. 
Therefore, as demonstrated in \citet{sb2015a}, a post-merger explosive gas expulsion is
instrumental in yielding YMCs that are like what we observe.   
On the other hand, if the sub-clusters are initially sufficiently apart that
gas expulsion occurs in them individually prior to the merger \citep{fk2005}, then
the likely outcome would be a highly diffuse, massive stellar association with substructures, and it may,
as a whole, be super-virial or sub-virial, \eg, Cyg-OB2.  

\section{Discussions and conclusions}\label{conclude}

While expansion of a star cluster, as driven by two-body relaxation,
binary heating and stellar mass loss,
is a fairly familiar phenomenon, its impact in shaping young star clusters remains to
be a matter of debate. The calculations in Sec.~\ref{purexp} show that if star
clusters preferably appear within the filamentary overdense structures of molecular clouds,
adapting to their typical dimensions of 0.1-0.3 pc, then the self-driven secular
expansion is generally insufficient for such clusters to reach the dimensions of observed YMCs.
This is true if the cluster forms monolithically or via hierarchical mergers of (closely located)
sub-clusters, and holds irrespective of the newborn cluster's mass. Note that
the monolithic cluster or the sub-clusters should naturally adapt to the compact sizes of the
filamentary and dense substructures inside molecular clouds.
While formation of proto-stars outside the filaments, \ie, in rarer media,
cannot be completely ruled out, both observations (\eg, \citealt{schn2010,schn2012,andr2013})
and hydrodynamic calculations (\eg, \citealt{kl1998,bate2004,giri2011})
suggest that the majority of the proto-stars must form within the compact
dimensions of the dense gas filaments. A fraction of them can thereafter
migrate to less dense parts of the local cloud via dynamical interactions.

The above direct N-body calculations naturally include an accurate treatment of
two-body relaxation and dynamical encounters involving binaries.
The other major processes that drive a cluster's expansion
are stellar mass loss (until $\approx50$ Myr) and dynamical heating by
compact remnants. As discussed in Sec.~\ref{purexp}, despite the
shortcomings of the implementation of stellar mass loss in {\tt NBODY7} (as such,
in any N-body code at present), the computed secular expansion is 
nearly independent of the details of stellar winds and supernovae and serves as an upper
limit. Also, while the computed models here begin with Plummer profiles,
such specific initial profile does not influence a cluster's overall expansion
rate, as determined by $\reff(t)$, which is driven by the overall stellar
mass loss and dynamical encounters in the inner part of the cluster.
In any case, a Plummer would be an appropriate initial
profile since the molecular clouds' filaments are found to possess
Plummer-like cross-sections \citep{mali2012}. In the present calculations including primordial
binaries, the massive O-star binaries, that are efficient in energy generation
through dynamical interactions, have a period distribution and binary
fraction that is consistent with observations (see Sec.~\ref{nbody}).     
This necessitates additional mechanisms that aid a young star cluster's
expansion.

Having run out of the other possibilities, expulsion of residual gas seems
to be the only process that evolves a newly hatched cluster from its compact
configuration to the observed sizes of YMCs, as the calculations of
Sec.~\ref{gasexp} show. However, the widest (and oldest) $\sim 10$ pc-sized
YMCs still seem to be difficult to achieve in this way (see Sec.~\ref{adexp}),
so that they are more likely to be ensembles of several young clusters (see Sec.~\ref{slowexp}). 

An important concern in the above line of argument is how YMCs and
SSCs gather their total mass of $\gtrsim10^4\Ms$ within a few Myr, irrespective
of their formation channels. This ``mass-budget problem'' 
in the context of cluster assembly is, at present, essentially unsettled,
both from observational and computational point of view. There is no clear
example of such an amount of assembling mass in molecular clouds in our Galaxy.
However, as discussed in the previous
sections, the assembly phase of starburst clusters is short, $\lesssim1$ Myr. Therefore,
it would anyway be rare to find a starburst cluster in its assembling phase
in a normal gas-rich disc galaxy like the Milky Way;
one finds either widely-distributed clouds or already-assembled VYMCs.
The chances of catching VYMCs and SSCs in their assembling phase would be higher
in starburst galaxies where a much larger number of massive clusters are triggered. Indeed,
recent ALMA observations of the Antennae Galaxy indicate still-forming, deeply-embedded
stellar systems of total mass exceeding $10^7\Ms$, which are either monolithic systems
or distributed over a few pc \citep{john2015a,john2015b}.
In order to better understand the conditions under which VYMCs and SSCs form,
improved and more exhaustive observations of starburst galaxies is necessary. 

Also, hydrodynamic simulations of molecular clouds cannot
presently be done with such a large mass,
keeping sufficient resolution at the same time. However,
a recent work by \citet{dale2015} involves low-resolution (mass resolution $10-100\Ms$)
smoothed-particle-hydrodynamics (SPH) calculations of massive ($10^5-10^6\Ms$),
turbulent molecular clouds. These calculations include radiative feedback onto the gas
from ionizing O/B stars but have no magnetic fields. A high-resolution (reaching
the ``opacity limit'') SPH computation over such a mass scale and also the inclusion
of magnetic field is technologically prohibitive at present. \citet{dale2015} find
sub-cluster formation at the deepest parts of the potential well where the SFE approaches
$\approx 100$\%. Furthermore, the gas is ionized and cleared from the neighborhood
of the sub-clusters,
causing essentially gas-free mergers of the sub-clusters. This inhibits
any mechanical effect of radiative gas blow out on the merged cluster, as \citet{dale2015} argue. 

Notably, these simulations are too poorly resolved to reliably infer any SFE. Also, 
since the sub-clusters themselves are treated as sink particles, it is impossible
to estimate the size of the final merged cluster.
The sink-particle treatment of sub-clusters would also grossly overestimate   
the SFE at their locations. High-resolution SPH
calculations (but for much less massive clouds, $\sim100\Ms$), including radiative feedback
and magnetic field (see Sec.~\ref{gasexp} and references therein), infer
SFE $\lesssim30$\%. Hence, although the most massive set of
molecular-gas SPH simulations to date, the treatment of gas in \citet{dale2015} and
its inferred role in assembling a cluster cannot be taken as conclusive. This
limitation is essentially imposed by the present technology.

The treatment of gas is simplistic in the present work as well. However,
it still captures the overall dynamical effect of gas blow-out on a newly-assembled
cluster. Note
that that a maximum localized SFE of 30\% is assumed in the computations
in Sec.~\ref{gasexp}, implying that in practice the impact of gas expulsion
could be stronger. Also, 
details like the (model) clusters' profiles and their spatial scales are
taken into account, on which the present arguments are based. The present
approach necessitates a significant role of rapid (explosive or near-explosive) gas blow out
that transforms infant clusters to observed profiles, in general. Notably,
with $\approx30$\% SFE, a $\approx10^5\Ms(10^4\Ms)$ cluster leaves $\approx80(30)$\%
of its mass bound (Brinkmann et al., in preparation), following an
explosive gas expulsion from compact initial conditions.
In other words, such gas expulsion does not
dissolve a YMC progenitor and retains a substantial fraction of it, in general
agreement with \citet{dale2015}. As pointed out in Secs.~\ref{intro} \& \ref{gasexp},
the efficient violent relaxation, due to the high stellar density in massive
compact proto-clusters, causes them to remain well bound (but to expand) even
after $\approx70$\% of their mass is expelled explosively.
Such a short gas-expulsion phase is, of course, difficult to catch
observationally. One such candidate is the RCW 38 cluster \citep{derose2009}, which
is highly compact ($\rh\approx0.1$ pc) and is exposed only over
its central region.

It may be argued that the observed YMCs are formed from $\rh(0)>1$ pc initial
configurations and with SFE $\approx100$\%, which would evolve to
the observed YMCs with $\reff\gtrsim1$ pc. From observations,
we also know that $\rh(0) \lesssim 0.3$ pc is common for infant clusters \citep{andr2011},  
implying that there should also exist a population
of gas-free clusters with ages $t>10$ Myr but $\reff\lesssim1$ pc, assuming again
an SFE $\approx100$\% (Fig.~\ref{fig:reffevol_noexp}), since in
more compact and dense environment the SFE is unlikely to decrease.
No such bimodal
population of exposed clusters is known, from which it follows that
$\rh(0)\lesssim0.3$ pc {\it and} SFE $\approx30$\% is the most
likely birth condition for YMCs. Also, the observed $\reff-t$ trend of the
present-day YMCs may be consistent with their expansion due to stellar
mass loss, with a considerable scatter though, \eg, as in \citet{ryon2015}.
However, this does not preclude an early gas-expulsion phase; once the
surviving cluster, after the gas depletion,
returns to (near) dynamical equilibrium \citep{sb2013},
the expanded cluster continues to expand (\cf Fig.~\ref{fig:reffevol_fastexp})
due to stellar mass loss and two-body relaxation processes (Sec.~\ref{secexp}). 

Notably, based on the velocity-space morphologies of gas clouds \citep{haworth2015} in the neighborhood of
several starburst clusters, some authors \citep{furuk2009,fukui2014,fukui2015}
infer that these clusters (\eg, Westerlund 2, NGC 3603) form out of
intense starbursts triggered during major cloud-cloud collisions. In this scenario,
the residual gas is effectively removed (without the aid of stellar feedback)
as the clouds cross and move away from each other. Such an interaction phase typically
lasts for $\lesssim1$ Myr, implying that the cluster has to form monolithically or
near monolithically \citep{sb2015a}. In this case, the cluster formed in the colliding
cloud would, as well, expand explosively, unless the conversion from gas to star
in the shocked region is nearly complete. As discussed in Sec.~\ref{gasexp}, both
observations and high-resolution SPH simulations, performed so far, seem to
disfavor the latter condition. A thorough survey of the kinematics
of gas around YMCs would help to reveal the role of the gas from which
these clusters were born.

\section*{Acknowledgements}

We thank the referee for constructive criticisms that led to substantial improvements of
the description and discussions in the paper.

\newpage

\onecolumn

\begin{longtable}{lllll}
\caption{\label{tab:cluster} A list of (stellar) age $t$, photometric mass $\mphot$ and effective radius
$\reff$ of bound star clusters, of age $t\lesssim100$ Myr (young) and photometric mass $\mphot\gtrsim10^4\Ms$
(massive), observed in the Milky Way Galaxy, the Local Group galaxies
and other external galaxies. This list is compiled from \citet{pz2010} (see their tables 2-4) where the references
for the data of the individual clusters are given. All these stellar systems are $\gtrsim10$ times
older than their present dynamical crossing time, $\tcr$, which is a phenomenological criterion to identify
them as gravitationally bound objects (see \citealt{pz2010}).}\\
\hline\hline
Galaxy name & Cluster name & Age $t$ (Myr) & $\logten(\mphot/\Ms)$ & $\reff$ (pc)\\
\hline
\endfirsthead
\caption{continued.}\\
Galaxy name & Cluster name & Age $t$ (Myr) & $\logten(\mphot/\Ms)$ & $\reff$ (pc)\\
\hline 
\endhead
\endfoot
Milky Way & Arches & 2.00 $(+2.00)$ &	4.30 &	0.40\\
Milky Way & DSB2003 & 3.50 &	3.80 &	1.20\\
Milky Way & NGC3603 & 2.00 $(-1.00)$ &	4.10 &	0.70\\
Milky Way & Quintuplet & 4.00 &	4.00 &	2.00\\
Milky Way & RSGC01 & 12.00 &	4.50 &	1.50\\
Milky Way & RSGC02 & 17.00 &	4.60 &	2.70\\
Milky Way & RSGC03 & 18.00 &	4.50 &	5.00\\
Milky Way & Trumpler14 & 2.00 &	4.00 &	0.50\\
Milky Way & Wd1 & 3.50 &	4.50 &	1.00\\
Milky Way & Wd2 & 2.00 &	4.00 &	0.80\\
Milky Way & hPer & 12.80 &	4.20 &	2.10\\
Milky Way & XPer & 12.80 &	4.10 &	2.50\\
\hline 
LMC  & R136 &  3.0 &	4.78 &	1.70\\
LMC  & NGC1818 & 25.1 &	4.42 &	5.39\\
LMC  & NGC1847 & 26.3 &	4.44 &	32.58\\
LMC  & NGC1850 & 31.6 &	4.86 &	11.25\\
LMC  & NGC2004 & 20.0 &	4.36 &	5.27\\
LMC  & NGC2100 & 15.8 &	4.36 &	4.41\\
LMC  & NGC2136 & 100.0 &	4.30 &	3.42\\
LMC  & NGC2157 & 39.8 &	4.31 &	5.39\\
LMC  & NGC2164 & 50.1 &	4.18 &	4.76\\
LMC  & NGC2214 & 39.8 &	4.03 &	8.13\\
LMC  & NGC1711 & 50.1 &	4.24 &	5.19\\
M31  & KW246  & 75.9 &	4.19 &	3.20\\
M31  & B257D  & 79.4 &	4.45 &	15.14\\
M31  & B318   & 70.8 &	4.38 &	6.61\\
M31  & B327   & 50.1 &	4.38 &	4.47\\
M31  & B448   & 79.4 &	4.58 &	16.22\\
M31  & Vdb0   & 25.1 &	4.85 &	7.40\\
M31  & KW044  & 58.9 &	4.59 &	10.00\\
M31  & KW120  & 87.1 &	4.57 &	2.60\\
M31  & KW208  & 56.2 &	4.01 &	2.90\\
M31  & KW272  & 53.7 &	4.50 &	9.00\\
M31  & B015D  & 70.8 &	4.76 &	16.60\\
M31  & B040   & 79.4 &	4.50 &	12.88\\
M31  & B043   & 79.4 &	4.43 &	3.98\\
M31  & B066   & 70.8 &	4.25 &	6.76\\
NGC6822  & HubbleIV & 25.1 &	4.00 &	2.00\\
SMC  & NGC330  &  25.1 &	4.56 &	6.11\\
\hline
ESO338IG & 23 & 7.08 &	6.70 &	5.20\\
M51 & 3cl-a   & 15.85 &	5.04 &	5.20\\
M51 & 3cl-b  & 5.01 &	5.91 &	2.30\\
M51 & a1  & 5.01 &	5.47 &	4.20\\
M82 & MGG9  & 9.55 &	5.92 &	2.60\\
M82 & A1 & 6.31 &	5.82 &	3.00\\
M82 & F & 60.26 &	6.70 &	2.80\\
NGC1140 & 1 & 5.01 &	6.04 &	8.00\\
NGC1487 & 2 & 8.51 &	5.20 &	1.20\\
NGC1487 & 1 & 8.32 &	5.18 &	2.30\\
NGC1487 & 3 & 8.51 &	4.88 &	2.10\\
NGC1569 & A & 12.02 &	6.20 &	2.30\\
NGC1569 & C & 3.02 &	5.16 &	2.90\\
NGC1569 & B & 19.95 &	5.74 &	2.10\\
NGC1569 & 30 & 91.20 &	5.55 &	2.50\\
NGC1705 & 1 & 15.85 &	5.90 &	1.60\\
NGC4038 & S2-1 & 8.91 &	5.47 &	3.70\\
NGC4038 & W99-1 & 8.13 &	5.86 &	3.60\\
NGC4038 & W99-16 & 10.00 &	5.46 &	6.00\\
NGC4038 & W99-2 & 6.61 &	6.42 &	8.00\\
NGC4038 & W99-15 &  8.71 &	5.70 &	1.40\\
NGC4038 & S1-1  & 7.94 &	5.85 &	3.60\\
NGC4038 & S1-2 & 8.32 &	5.70 &	3.60\\
NGC4038 & S1-5 & 8.51 &	5.48 &	0.90\\
NGC4038 & 2000-1 &  8.51 &	6.23 &	3.60\\
NGC4038 & S2-2 & 8.91 &	5.60 &	2.50\\
NGC4038 & S2-3 & 8.91 &	5.38 &	3.00\\
NGC4449 & N-1 & 10.96 &	6.57 &	16.90\\
NGC4449 & N-2 & 3.02 &	5.00 &	5.80\\
NGC5236 & 805 & 12.59 &	5.29 &	2.80\\
NGC5236 & 502 & 100.00 & 5.65 &	7.60\\
NGC5253 & I & 11.48 &	5.38 &	4.00\\
NGC5253 & VI & 10.96 &	4.93 &	3.10\\
NGC6946 & 1447 & 11.22 &	5.64 &	10.00\\
\hline\hline
\end{longtable}

\begin{longtable}{lllll}
\caption{\label{tab:assoc} A list of (stellar) age $t$, photometric mass $\mphot$ and effective radius
$\reff$ of young, massive (see Table~\ref{tab:cluster}; text)
stellar associations observed in the Milky Way Galaxy, the Local Group of galaxies
and other external galaxies. This list is compiled from \citet{pz2010} (see their tables 2-4) where the references
for the data of the individual clusters are given. All these stellar systems are
$\lesssim5$ times younger than their present dynamical crossing time,
$\tcr$ (\ie, they are dynamically young), which implies that they may or may
not be gravitationally bound objects.}\\
\hline\hline
Galaxy name & Cluster name & Age $t$ (Myr) & $\logten(\mphot/\Ms)$ & $\reff$ (pc)\\
\hline
\endfirsthead
\caption{continued.}\\
Galaxy name & Cluster name & Age $t$ (Myr) & $\logten(\mphot/\Ms)$ & $\reff$ (pc)\\
\hline 
\endhead
\endfoot
Milky Way & CYgOB & 2.5 &	4.40 &	5.20\\
Milky Way & IC1805 & 2.00 &	4.20 &	12.50\\
Milky Way & ILac1 & 14.00 &	3.40 &	20.70\\
Milky Way & LowerCenCrux & 11.50 &	3.30 &	15.00\\
Milky Way & NGC2244 & 2.00 &	3.90 &	5.60\\
Milky Way & NGC6611 & 3.00 &	4.40 &	5.90\\
Milky Way & NGC7380 & 2.00 &	3.80 &	6.50\\
Milky Way & ONC & 1.00 &	3.65 &	2.00\\
Milky Way & OriIa & 11.40 &	3.70 &	16.60\\
Milky Way & OriIb & 1.70 &	3.60 &	6.30\\
Milky Way & OriIc & 4.60 &	3.80 &	12.50\\
Milky Way & UpperCenCrux & 14.50 &	3.60 &	22.10\\
Milky Way & USco & 5.50 &	3.50 &	14.20\\
\hline
M31 & KW249  & 5.0 &	4.30 &	13.50\\
M31 & KW258  & 5.0 &	4.05 &	3.40\\
M33 & NGC595 & 4.0 &	4.50 &	26.90\\
M33 & NGC604 & 3.5 &	5.00 &	28.40\\
SMC & NGC346 & 3.0 &	5.60 &	9.00\\
\hline
NGC2403 & I-B & 6.03 &	4.82 &	26.30\\
NGC2403 & I-C & 6.03 &	4.42 &	19.60\\
NGC2403 & I-A & 6.03 &	5.06 &	20.60\\
NGC2403 & II & 4.47 &	5.35 &	11.80\\
NGC2403 & IV & 4.47 &	5.07 &	30.00\\
NGC4214 & VI & 10.96 &	4.93 &	35.90\\
NGC4214 & V & 10.96 &	5.73 &	83.90\\
NGC4214 & VII &  10.96 &	5.33 &	40.40\\
NGC4214 & I-A & 3.47 &	5.44 &	16.50\\
NGC4214 & I-B & 3.47 &	5.40 &	33.00\\
NGC4214 & I-D &  8.91 &	5.30 &	15.30\\
NGC4214 & II-C & 2.00 &	4.86 &	21.70\\
NGC5253 & IV & 3.47 &	4.72 &	13.80\\
\hline\hline
\end{longtable}

\begin{longtable}{rccccccc}
\caption{\label{tab:complist_nogas} Initial conditions for the computed model clusters
without a gas expulsion phase (see Sec.~\ref{nbody}).
The initial configurations
are Plummer profiles with total mass, $\mcl(0)$, and half-mass radius, $\rh(0)$.
The corresponding values of the virial velocity dispersion
(velocity dispersion within the virial radius $\approx\rh(0)$), $V_\ast(0)$,
the crossing time, $\tcr(0)=\rh(0)/V_\ast(0)$,
and the two-body relaxation time at half-mass radius, $\tauh(0)$, are given.
It is also indicated if the initial system
includes primordial binaries (see Sec.~\ref{nbody}) and if it is mass segregated
(as in \citealt{bg2008}).
Stellar evolution (see Sec.~\ref{nbody})
is used in all N-body calculations. The initial clusters are always located at
$R_G\approx8$ kpc Galactocentric distance (the solar distance)
and they orbit the Galactic center with $V_G\approx220\kmps$.}\\
\hline\hline
$\mcl(0)/\Ms$ & $\rh(0)/{\rm pc}$ & Primordial binaries & Stellar evolution & Mass segregation & $V_\ast(0)/\kmps$ & $\tcr(0)/{\rm Myr}$ & $\tauh(0)/{\rm Myr}$\\  
\hline
\endfirsthead
\caption{continued.}\\
$\mcl(0)/\Ms$ & $\rh(0)/{\rm pc}$ & Primordial binaries & Stellar evolution & Mass segregation & $V_\ast(0)/\kmps$ & $\tcr(0)/{\rm Myr}$ & $\tauh(0)/{\rm Myr}$\\  
\hline
\endhead
\endfoot
$10^4$ & 0.3 & no & yes & no & 10.5 & 0.028 & 4.65 \\ 
$3\times10^4$ & 0.3 & no & yes & no & 18.1 & 0.017 & 8.05 \\ 
$10^4$ & 1.0 & no & yes & no & 5.8 & 0.172 & 28.30 \\ 
$3\times10^4$ & 1.0 & no & yes & no & 10.0 & 0.100 & 49.02 \\ 
$10^5$ & 1.0 & no & yes & no & 18.2 & 0.055 & 89.49 \\ 
$1.3\times10^4$ & 0.35 & yes & yes & yes & 11.7 & 0.030 & 6.68 \\ 
$10^4$ & 1.0 & yes & yes & yes & 6.0 & 0.167 & 28.30 \\ 
$3\times10^4$ & 1.0 & yes & yes & yes & 10.4 & 0.096 & 49.02 \\ 
\hline\hline
\end{longtable}

\begin{longtable}{rcrccccc}
\caption{\label{tab:complist_gexp} Initial conditions and gas expulsion parameters
for the computed model clusters that include a gas expulsion
phase. The columns from left to right give (a) total initial stellar mass, $\mcl(0)$,
(b) initial stellar half-mass radius, $\rh(0)$,
(c) initial gas (potential, see Sec.~\ref{gasexp}) mass, $\mg(0)$,
that follows the stellar profile, (d) virial velocity dispersion, $V_\ast(0)$,
(e) scaled velocity dispersion, $V^\prime_\ast(0)$, (f) the corresponding
crossing time (dynamical time), $\tcr^\prime(0)$,
(g) timescale for gas (potential) dilution, $\tg$, and (h) the time,
$\td$, at which the gas expulsion begins. Here, $V^\prime_\ast(0)$
is the virial velocity dispersion at which the stellar velocities are boosted
so that the intended $\rh(0)$ can be maintained
when the gas potential is applied. For $\mg(0)\approx2\mcl(0)$
(\ie, $\epsilon\approx33$\% star formation efficiency), as in here,
$V^\prime_\ast(0)\approx\sqrt3V_\ast(0)$, which determines the
dynamical time, $\tcr^\prime(0)$, of the initial gas-embedded clusters.
These initial clusters are located at $R_G\approx8$ kpc Galactocentric distance (the solar distance)
and they orbit the Galactic center with $V_G\approx220\kmps$.
}\\
\hline\hline
$\mcl(0)/\Ms$ & $\rh(0)/{\rm pc}$ & $\mg(0)/\Ms$ & $V_\ast(0)/\kmps$ & $V^\prime_\ast(0)/\kmps$ & $\tcr^\prime(0)/{\rm Myr}$ & $\tg/{\rm Myr}$ & $\td/{\rm Myr}$\\  
\hline
\endfirsthead
\caption{continued.}\\
$\mcl(0)/\Ms$ & $\rh(0)/{\rm pc}$ & $\mg(0)/\Ms$ & $V_\ast(0)/\kmps$ & $V^\prime_\ast(0)/\kmps$ & $\tcr^\prime(0)/{\rm Myr}$ & $\tg/{\rm Myr}$ & $\td/{\rm Myr}$\\  
\hline
\endhead
\endfoot
$10^4$ & 0.15 & $2\times10^4$ & 14.7 & 25.5 & 0.006 & 0.015 & 0.6\\  
$3\times10^4$ & 0.15 & $6\times10^4$ & 25.8 & 44.7 & 0.003 & 0.015 & 0.6\\
$10^4$ & 0.30 & $2\times10^4$ & 10.5 & 18.2 & 0.016 & 0.030 & 0.6\\
$3\times10^4$ & 0.30 & $6\times10^4$ & 18.1 & 31.3 & 0.010 & 0.030 & 0.6\\  
$10^5$ & 0.30 & $2\times10^5$ & 33.1 & 57.3 & 0.005 & 0.030 & 0.6\\
$10^4$ & 1.0 & $2\times10^4$ & 5.8 & 10.0 & 0.100 & 0.100 & 0.6\\
$3\times10^4$ & 1.0 & $6\times10^4$ & 10.0 & 17.3 & 0.058 & 0.100 & 0.6\\
$10^5$ & 1.0 & $2\times10^5$ & 18.2 & 31.5 & 0.032 & 0.100 & 0.6\\
\hline
$3\times10^4$ & 1.0 & $6\times10^4$ & 9.9 & 17.1 & 0.058 & 2.5, 5.0 & 0.0 \\
$3\times10^4$ & 2.0 & $6\times10^4$ & 7.0 & 12.1 & 0.165 & 2.5, 5.0 & 0.0 \\
$3\times10^4$ & 3.0 & $6\times10^4$ & 5.8 & 10.0 & 0.300 & 2.5, 5.0 & 0.0 \\
$3\times10^4$ & 4.0 & $6\times10^4$ & 5.0 & 8.7  & 0.460 & 2.5, 5.0 & 0.0 \\
$10^5$ & 2.0 & $2\times10^5$ & 12.9 & 22.3 & 0.090 & 5.0 & 0.0\\ 
$10^5$ & 3.0 & $2\times10^5$ & 10.5 & 18.2 & 0.165 & 5.0 & 0.0\\ 
$10^5$ & 4.0 & $2\times10^5$ & 9.1 & 15.8  & 0.253 & 5.0 & 0.0\\
\hline\hline
\end{longtable}

\begin{thebibliography}{}

\bibitem[Aarseth(2003)]{aseth2003}
Aarseth, S.J. 2003, ``Gravitational N-Body Simulations''. Cambridge University Press.

\bibitem[Aarseth(2012)]{aseth2012}
Aarseth, S.J. 2012, 422, 841.

\bibitem[Amaro-Seoane et al.(2014)]{ams2014}
Amaro-Seoane, P., Konstantinidis, S., Freitag, M.D., et al. 2014, \apj, 782, 97.

\bibitem[Andr\'e et al.(2011)]{andr2011}
Andr\'e, P., Me\'nshchikov, A., Koenyves, V., et al. 2011,
in Alfaro Navarro, E.J., Gallego Calvente, A.T., Zapatero Osorio, M.R. (Eds.)
\emph{Stellar Clusters \& Associations: A RIA Workshop on Gaia}. 
Granada, Spain: IAA-CSIC, 321.

\bibitem[Andr\'e et al.(2014)]{andr2013}
Andr\'e, P., Di Francesco, J., Ward-Thompson, D., et al. 2014,
in Beuther, H., Klessen, R., Dullemond, C. and Henning, Th. (Eds.)
\emph{Protostars and Planets VI}, University of Arizona Press, Tucson, p.27.

\bibitem[Bally et al.(2012)]{bally2012}
Bally, J., Walawender, J. and Reipurth, B. 2012, \aj, 144, 143.

\bibitem[Bally et al.(2015)]{bally2015}
Bally, J., Ginsburg, A., Silvia, D. and Youngblood, A. 2015, \aap, 579, id.A130.

\bibitem[Banerjee et al.(2010)]{sb2010}
Banerjee, S., Baumgardt, H. and Kroupa, P. 2010, \mnras, 402, 371.

\bibitem[Banerjee(2011)]{sb2011}
Banerjee, S. 2011 in ``Proceedings of the 25th Texas Symposium on
Relativistic Astrophysics - TEXAS 2010'' (Proceedings of Science),
arXiv:1102.4614 (preprint).    

\bibitem[Banerjee \& Kroupa(2012)]{sb2012}
Banerjee, S. and Kroupa, P. 2012, \aap, 547, id.A23.

\bibitem[Banerjee et al.(2012)]{sbetl2012}
Banerjee, S., Kroupa, P. and Oh. S. 2012, \apj, 746, 15.

\bibitem[Banerjee \& Kroupa(2013)]{sb2013}
Banerjee, S. and Kroupa, P. 2013, \apj, 764, 29.

\bibitem[Banerjee \& Kroupa(2014)]{sb2014}
Banerjee, S. and Kroupa, P. 2014, \apj, 787, 158.

\bibitem[Banerjee \& Kroupa(2015a)]{sb2015a}
Banerjee, S. and Kroupa, P. 2015a, \mnras, 447, 728.

\bibitem[Banerjee \& Kroupa(2015b)]{sb2015b}
Banerjee, S. and Kroupa, P. 2015b, in
S.W. Stahler eds. ``The Birth of Star Clusters'', Springer-Verlag, preprint URL:\\
\url{http://tinyurl.com/op7zeb6/docs/VYMCchapter.pdf}

\bibitem[Bastian \& Silva-Villa(2013)]{bsv2013}
Bastian, N. and Silva-Villa, E. 2013, \mnras, 431, L122. 

\bibitem[Bate \& Bonnell(2004)]{bate2004}
Bate M.R. and Bonnell, I.A. 2004, in Lamers, H.J.G.L.M., Smith, L.J., Nota A. (Eds.)
\emph{The Formation and Evolution of Massive Young Star Clusters},
(ASP Conf. Proc. 322). San Francisco: Astronomical Society of the Pacific, 289.

\bibitem[Bate(2009)]{bate2009}
Bate M.R., 2009, \mnras, 392, 590.

\bibitem[Bate(2012)]{bate2012}
Bate, M.R. 2012, \mnras, 419, 3115.

\bibitem[Bate et al.(2014)]{bate2013}
Bate, M.R., Tricco, T.S., Price, D.J. 2014, \mnras, 437, 77.

\bibitem[Baumgardt \& Kroupa(2007)]{bk2007}
Baumgardt, H. and Kroupa, P., 2007, \mnras, 380, 1589.

\bibitem[Baumgardt et al.(2008)]{bg2008}
Baumgardt, H., De Marchi, G. and Kroupa, P. 2008, \apj, 685, 247.

\bibitem[Bik et al.(2014)]{bik2014}
Bik, A., Stolte, A., Gennaro, M., et al. 2014, \aap, 561, id.A12.

\bibitem[Boily \& Kroupa(2003a)]{bk2003a}
Boily, C. and Kroupa, P. 2003a, 338, 665.

\bibitem[Boily \& Kroupa(2003b)]{bk2003b}
Boily, C. and Kroupa, P. 2003b, 338, 673.

\bibitem[Br\"uns et al.(2009)]{bru2009}
Br\"uns, R.C., Kroupa, P., Fellhauer, M. 2009, \apj, 702, 1268.

\bibitem[Dale et al.(2015)]{dale2015}
Dale, J.E., Ercolano, B. and Bonnell, I.A. 2015, \mnras, 451, 5506.

\bibitem[de Mink et al.(2009)]{dmink2009}
de Mink, S.E., Pols, O.R., Langer, N., Izzard, R.G. 2009, \aap, 507, L1. 

\bibitem[DeRose et al.(2009)]{derose2009}
DeRose, K.L., Bourke, T.L., Gutermuth, R.A., et al. 2009, \aj, 138, 33.

\bibitem[Duarte-Cabral et al.(2011)]{duca2011}
Duarte-Cabral, A., Dobbs, C.L., Peretto, N., et al. 2011, \aap, 528, A50.  

\bibitem[Elmegreen(1983)]{elmg1983}
Elmegreen, B.G. 1983, \mnras, 203, 1011.

\bibitem[Feigelson \& Townsley(2008)]{fglsn2008}
Feigelson, E.D. and Townsley, L.K. 2008, \apj, 673, 354.

\bibitem[Fellhauer \& Kroupa(2005)]{fk2005}
Fellhauer, M. and Kroupa, P. 2005, \apj, 630, 879. 

\bibitem[Fujii et al.(2012)]{fuji2012}
Fujii, M.S., Saitoh, T.R. and Portegies Zwart, S.F. 2012, \apj, 753, 85.

\bibitem[Fukui et al.(2014)]{fukui2014}
Fukui, Y., Ohama, A., Hanaoka, N., et al. 2014, \apj, 780, 36. 

\bibitem[Fukui et al.(2015)]{fukui2015}
Fukui, Y., Torii, K., Ohama, A., et al. 2015, \apj, arXiv:1504.05391 (preprint).

\bibitem[Furukawa et al.(2009)]{furuk2009}
Furukawa, N., Dawson, J.R., Ohama, A., et al. 2009, \apj, 696, L11. 

\bibitem[Gieles et al.(2012)]{gieles2012}
Gieles, M., Moeckel N. and Clarke, C.J. 2012, \mnras, 426, L11.

\bibitem[Girichidis et al.(2011)]{giri2011}
Girichidis, P., Federrath, C., Banerjee, R. and Klessen, R.S. 2011, \mnras,  
413, 2741.

\bibitem[Girichidis et al.(2012)]{giri2012}
Girichidis, P., Federrath, C., Banerjee, R. and Klessen, R.S. 2012, \mnras,  
420, 613.

\bibitem[Gratton et al.(2012)]{grat2012}
Gratton, R.G., Carretta, E. and Bragaglia, A. 2012, A\&ARv, 20, 50. 

\bibitem[Haworth et al.(2015)]{haworth2015}
Haworth, T.J., Tasker, E.J., Fukui, Y., et al. 2015, \mnras, 450, 10.

\bibitem[Heggie(1975)]{hh1975}
Heggie, D.C. 1975, \mnras, 173, 729.

\bibitem[Heggie \& Hut(2003)]{hh2003} 
Heggie, D.C. and Hut, P. 2003, ``The Gravitational Millon-Body Problem: A Multidisciplinary Approach to
Star Cluster Dynamics''. Cambridge University Press, Cambridge, UK.

\bibitem[H\`enon(1965)]{hen1965}
H\`enon, M. 1965, {\it Ann. Astrophys.}, 28, 62.

\bibitem[Hollyhead et al.(2015)]{holly2015}
Hollyhead, K., Bastian, N., Adamo, A., et al. 2015, \mnras, 449, 1106. 

\bibitem[Hurley et al.(2000)]{hur2000}
Hurley, J.R., Pols, O.R. and Tout, C.A. 2000, \mnras, 315, 543.

\bibitem[Hurley et al.(2002)]{hur2002}
Hurley, J.R., Tout, C.A. and Pols, O.R. 2002, \mnras, 329, 897.

\bibitem[Johnson(2015)]{john2015a}
Johnson, K. 2015, IAU GA Meeting 29, id.2257770.

\bibitem[Johnson et al.(2015)]{john2015b}
Johnson, K.E., Leroy, A.K., Indebetouw, R., et al. 2015, \apj, 806, 35.

\bibitem[Klessen et al.(1998)]{kl1998}
Klessen, R.S., Burkert, A. and Bate, M.R. 1998, \apjl, 501, L205.

\bibitem[Kroupa(1995)]{pk1995b}
Kroupa, P. 1995, \mnras, 277, 1507.

\bibitem[Kroupa(2001)]{pk2001}
Kroupa, P. 2001, \mnras, 322, 231.

\bibitem[Kroupa et al.(2001)]{pketl2001}
Kroupa, P., Aarseth, S. and Hurley, J. 2001, \mnras, 321, 699.

\bibitem[Kroupa(2008)]{pk2008}
Kroupa, P. 2008, in Aarseth, S.J., Tout, C.A., Mardling, R.A., eds,
Lecture Notes in Physics Vol. 760, Initial Conditions for Star Clusters.
Springer-Verlag, Berlin, p. 181.

\bibitem[Kruijssen et al.(2012)]{kru2012}
Kruijssen, J.M.D., Maschberger, T., Moeckel, N., et al. 2012, \mnras, 419, 841. 

\bibitem[Lada \& Lada(2003)]{lnl2003}
Lada, C.J. and  Lada, E.A. 2003, \araa, 41, 57.

\bibitem[Longmore et al.(2014)]{longm2014}
Longmore, S.N., Kruijssen, J.M.D., Bastian, N., et al. 2014,
in Beuther, H., Klessen, R., Dullemond, C. and Henning, Th. (Eds.)
\emph{Protostars and Planets VI}, University of Arizona Press, Tucson, p.291.

\bibitem[Machida \& Matsumoto(2012)]{mnm2012}
Machida, M.N. and Matsumoto, T. 2012, \mnras, 421, 588.

\bibitem[Mackey \& Gilmore(2003a)]{macgil2003a}
Mackey, A.D. and Gilmore, G.F. 2003a, \mnras, 338, 85.

\bibitem[Mackey \& Gilmore(2003b)]{macgil2003b}
Mackey, A.D. and Gilmore, G.F. 2003b, \mnras, 338, 120.

\bibitem[Mackey et al.(2008)]{macetl2008}
Mackey, A.D., Wilkinson, M.I., Davies, M.B. and Gilmore, G.F. 2008, \mnras, 386, 65. 

\bibitem[Malinen et al.(2012)]{mali2012}
Malinen, J., Juvela, M., Rawlings, M.G., et al. 2012, \aap, 544, id.A50.

\bibitem[Marks \& Kroupa(2012)]{mk2012}
Marks, M. and Kroupa, P. 2012, 543, A8.

\bibitem[Massi et al.(2015)]{massi2014}
Massi, F., Giannetti, A., di Carlo, E. 2015, \aap, 573, id.A95. 

\bibitem[Messineo et al.(2015)]{messi2015}
Messineo, M., Clark, J.S., Figer, D.F., et al. 2015, \apj, 805, 110.

\bibitem[Morscher et al.(2013)]{morscher2013}
Morscher, M., Umbreit, S., Farr, W.M. and Rasio, F.A. 2013, \apj, 763, L15.

\bibitem[Oh et al.(2015)]{oh2015}
Oh, S., Kroupa, P. and Pflamm-Altenburg, J. 2015, \apj, 805, 92.

\bibitem[Pfalzner(2009)]{pf2009}
Pfalzner, S. 2009, \aap, 498, L37.

\bibitem[Pfalzner \& Kaczmarek(2013)]{pfkz2013}
Pfalzner, S. and Kaczmarek, T. 2013, \aap, 559, A38.

\bibitem[Pfalzner et al.(2014)]{pf2014}
Pfalzner, S., Parmentier, G., Steinhausen, M., et al. 2014, \apj, 794, 147. 

\bibitem[Pflamm-Altenburg \& Kroupa(2009)]{pflm2009}
Pflamm-Altenburg, J. and Kroupa, P. 2009, \mnras, 397, 488.

\bibitem[Plummer(1911)]{plum1911}
Plummer, H.C. 1911, \mnras, 71, 460.

\bibitem[Portegies Zwart et al.(2010)]{pz2010}
Portegies Zwart, S.F., McMillan, S.L.W. and Gieles, M. 2010, \araa, 48, 431.

\bibitem[Rathborne et al.(2015)]{rath2015}
Rathborne, J.M., Longmore, S.N., Jackson, J.M., et al. 2015, \apj, 802, 125.

\bibitem[Renaud et al.(2015)]{renaud2015}
Renaud, F., Bournaud, F. and Duc, P-A. 2015, \mnras, 446, 2038.

\bibitem[Rom\'an-Z\'uniga et al.(2015)]{rz2015}
Rom\'an-Z\'uniga, C. G., Ybarra, J., Megias, G., et al. 2015, \apj, arXiv:1507.00016 (preprint).

\bibitem[Ryon et al.(2015)]{ryon2015}
Ryon, J.E., Bastian, N., Adamo, A., et al. 2015, \mnras, arXiv:1506.02042 (preprint).

\bibitem[Sana \& Evans(2011)]{sev2011}
Sana, H. and Evans, C.J. 2011,
in Neiner, C., Wade, G., Meynet, G. and Peters, G. (Eds.)
\emph{Active OB Stars: Structure, Evolution, Mass Loss, and Critical Limits} (IAU Symp. 272).
Cambridge Univ. Press, Cambridge, 474.

\bibitem[Schneider et al.(2010)]{schn2010}
Schneider, N., Csengeri, T., Bontemps, S., et al. 2010, \aap, 520, A49.

\bibitem[Schneider et al.(2012)]{schn2012}
Schneider, N., Csengeri, T., Hennemann, M., et al. 2012, \aap, 540, L11.

\bibitem[Smith et al.(2013)]{sm2013}
Smith, R., Goodwin, S., Fellhauer, M. and Assmann, P. 2013, \mnras, 428, 1303.

\bibitem[Spitzer(1987)]{spitz87}
Spitzer, L. Jr. 1987, ``Dynamical Evolution of Globular Clusters'',
Princeton University Press.

\bibitem[Tafalla \& Hacar(2014)]{tafhac2015}
Tafalla, M. and Hacar, A. 2014, \aap, 574, id.A104.

\bibitem[Tapia et al.(2011)]{tapia2011}
Tapia, M., Roth, M., Bohigas, J., et al. 2011, \mnras, 416, 2163. 

\bibitem[Traficante et al.(2015)]{traf2015}
Traficante, A., Fuller, G.A., Peretto, N., et al. 2015, \mnras, 451, 3089.

\bibitem[Weidner \& Kroupa(2004)]{wk2004}
Weidner, C. and Kroupa, P. 2004, \mnras, 348, 187.

\bibitem[Weidner et al.(2013)]{wetl2013}
Weidner, C., Kroupa, P. and Pflamm-Altenburg, J. 2013, \mnras, 434, 84.

\end{thebibliography}
\end{document}